\documentclass [12pt]{article}
\usepackage{amsmath,amssymb,amsthm}
\usepackage{color}
\usepackage{multirow} 
\newtheorem{proposition}{Proposition}
\usepackage{amssymb,amsthm}
\usepackage{graphicx}

\newtheorem{theorem}{Theorem}
\newtheorem{corollary}{Corollary}
\newtheorem{definition}{Definition}
\newtheorem{lemma}{Lemma}

\newcommand{\norm}[1]{\lVert#1\rVert}
\newcommand{\ip}[2]{\langle#1,#2\rangle}

\def\N{\mathbb{N}}
\def\R{\mathbb{R}}

\def\C^4{\mathcal{L}}

\def\a{\alpha}
\def\a{\alpha}

\def\F{\mathcal F}

\def\rad{\textrm{rad}}

\def\vol{\textrm{ vol}}
\def\area{\textrm{ area}}

\def\db{\textcolor{black}}

\begin{document}
\title{Region-of-Interest reconstruction from truncated cone-beam projections}

\author{Robert~Azencott${}^1$, Bernhard~G. Bodmann${}^1$, Tasadduk~Chowdhury${}^1$, \\Demetrio~Labate${}^1$, Anando~Sen${}^2$, Daniel Vera${}^3$
}

\footnotetext[1]{Department of Mathematics, University of Houston, Houston, TX 77204, USA.}
\footnotetext[2]{Department of Biomedical Informatics, Columbia University, New York, NY, USA.}
\footnotetext[3]{Matematicas, Instituto Tecnologico Autonomo de Mexico, Mexico}
\date{}

\maketitle

\begin{abstract}
Region-of-Interest (ROI) tomography aims at reconstructing a region of interest $C$ inside a body using only  x-ray projections intersecting $C$ with the goal to reduce overall radiation exposure when only a small specific region of the body needs to be examined. We consider x-ray acquisition from  sources located on a smooth curve $\Gamma$ in $\R^3$ verifying classical Tuy's condition.  In this situation,  the  {\it non-trucated} cone-beam transform $D f$ of smooth densities $f$ admits an explicit  inverse $Z$; however $Z$ cannot directly reconstruct $f$ from ROI-truncated projections. To deal with the ROI tomography problem, we introduce a novel reconstruction approach. For densities $f$ in $L^{\infty}(B)$ where $B$ is a bounded ball in $\R^3$, our method iterates  an operator $U$ combining  ROI-truncated projections, inversion by the  operator $Z$ and appropriate regularization operators. Assuming only knowledge of projections corresponding to a spherical ROI $C \subset B$, given $\epsilon >0$, we prove that if $C$ is sufficiently large our iterative reconstruction algorithm converges uniformly to an $\epsilon$-accurate approximation of $f$, where the accuracy depends on the regularity of $f$ quantified in the Sobolev norm $W^5(B)$.  This result shows the existence of a critical ROI radius ensuring  the convergence of the ROI reconstruction algorithm to $\epsilon$-accurate approximations of $f$. We numerically verified these theoretical results using simulated acquisition of ROI-truncated cone-beam projection data for multiple acquisition geometries. Numerical experiments indicate that the critical ROI radius is fairly small with respect to the support region~$B$.

\end{abstract}



\section{Introduction}

 Computed Tomography (CT) is a non-invasive imaging technique, routinely used in medical diagnostics and interventional surgical procedures to visualize specific regions inside a body. CT involves patient exposure to x-ray
radiation, with health risks of radiation-induced carcinogenesis
which are essentially proportional to radiation exposure levels
~\cite{lee:dose, huda:embryo}. To reduce radiation exposure in CT,
several strategies have been explored such as sparsifying the
numbers of x-ray projections or truncating the projections so that
only x-rays intersecting a small region-of-interest (ROI)
are acquired. 
Reconstructing a density $f$ from its  projections is an ill-posed problem, meaning that small perturbations of the projections may lead to significant reconstruction errors. To address this problem, several approximate or regularized reconstruction formulas have been introduced over the years, such as the classical Filtered Back-Projection or the FDK algorithms \cite[Ch.5]{natterer:imagerec}. However these methods are designed to work using non-truncated   projection data. When  projections are truncated, the reconstruction problem may become severely ill posed \db{and non-uniquely solvable~\cite{natterer:tomography}. For instance, the so-called {\it interior problem}, where projection data are only known on a region strictly inside the support of the density $f$, has no unique solution in general \cite{natterer:tomography}. As a result,} naive numerical reconstruction algorithms such as direct application of a global reconstruction formula, with the missing projection data set to zero, typically produce serious instability and unacceptable visual artifacts.

\paragraph{The ROI reconstruction problem.} The problem of ROI recontruction in CT has been studied in multiple papers and using a variety of methods (see, for example, the recent reviews~\cite{clackdoyle:roi,WangYu13} and the references therein).  Recent remarkable results have shown that it is often possible to derive analytic ROI reconstruction formulas from truncated projections, provided the ROI is chosen with certain restrictions  (cf.~\cite{noo:imrec,clackdoyle:alcif, zou:image}). Such explicit ROI reconstruction formulas from truncated projections typically depend on the specific acquisition modalities and impose restrictions on ROI geometry;  for instance, some prior partial knowledge of the density $f$ within the ROI is required or the ROI cannot lie strictly inside the support of $f$. 

\db{
Iterative methods on the other hand provide a more flexible alternative for the reconstruction from truncated or incomplete projections as they can be applied to essentially any type of acquisition mode (cf.~\cite{herman:sparse, sidky:sparse, zeng:iterative}). Many such methods rely on total variation  and other forms of regularization to ensure the convergence of the algorithm. For instance, the recent ROI reconstruction approach by Klann et al. \cite{Klann2015} relies on an appropriate wavelet based regularization. In this approach, the uniqueness of the interior problem is guaranteed under the hypothesis that the density function is piecewise constant. However this result assumes the ideal case of a noiseless acquisition and leaves the problem of stability in the presence of noise open. With respect to analytic formulas,} iterative methods are usually computationally more intensive, especially for 3D data. However, advances in computational capabilities (e.g., \cite{Guorui:katsevichGPU}) and recent ideas from compressed sensing (e.g., \cite{YuWang09}) offer powerful tools to overcome this limitation. 

\paragraph{Our approach.}
\db{
In this paper, we consider the {\it ROI reconstruction problem} aiming at reconstructing an unknown density inside an ROI $C$ using only the projections intersecting $C$.  We fix a bounded ball $B \subset \R^3$ and a smooth curve $\Gamma \subset \R^3$ of `ray sources' exterior to $B$. We assume that the objects illuminated by x-rays are strictly included in $B$ and are characterized by their  unknown density functions $f \in L^{\infty}(B)$. For any spherical region of interest $C \subset B$, we denote by  $\mathcal{R}_C$ the set of all half-lines (or  `half-rays') $r$ emanating from arbitrary points  of $\Gamma$ and  intersecting $C$. The $C$-truncated cone beam projection operator $D_C$ maps any density function $f \in L^{\infty}(B)$ into a function $D_C f$ defined, for each half-ray $r \in \mathcal{R}_C$, by integrating $f$ over~$r$. }

\db{
The  {\it non truncated} cone beam  operator, corresponding to the case $C=B$, will be denoted $D = D_B$. It is known that, if  $\Gamma$ and $B$ verify classical geometric  {\it Tuy's condition}, Grangeat's classical formula provides an inverse $Z$ of the non truncated cone beam projection $D$, verifying $Z D f = f $ for every $C^2$-density $f$ with compact support included in $B$. For several specific types of curves $\Gamma$, the non truncated operator  $D$ can classically  be  inverted  by an operator $Z$ defined on smooth densities by  geometry-specific formulas.} 

\db{
In the setting described above, we develop a method to construct approximate inverses $Z_C$  for the $C$-truncated  operator $D_C$, defined by an iterative  $ROI$-reconstruction algorithm  which converges whenever the unknown density $f$  is smooth enough and the volume of the difference set $B \setminus C$ is small enough.
More precisely, we start with  an  explicit operator $Z$ inverting the {\it non-truncated} cone beam projection  $D$ for  smooth densities and we construct   a regularization operator $\tau$ such that  $U = Z \tau (D - D_C)$ becomes a contraction on $L^\infty(B)$. Then, for  any unknown density $f$ in $W^5(B)$, we set $f_0= Z \tau D_C f$  to define iteratively the  approximating sequence of densities $(f_j)$ by  
\begin{equation} \label{def.algo} 
f_{j+1} = f_0 +U f_j, 
\end{equation}
Our mathematical analysis proves that as $j \to \infty$, the sequence  $f_j$  converges to an $\epsilon$-accurate reconstruction $\hat{f}$ of $f$, at exponential speed in $L^{\infty}(C)$, for any spherical ROI $C \subset B$ having a radius larger than a {\it critical radius} $ \rho(\epsilon)$. That is, given an accuracy level $\epsilon$ and a sufficiently large spherical region $C \subset B$, we generate an estimate $\hat{f} $ of $f$ such that 
$$
\norm{ \hat{f} - f }_{ L^{\infty}(C)} \leq \epsilon \norm{f}_{ W^5(B)}.
$$
Our results also extend to the situation where sources are located over a whole sphere in $\R^3$ containing $B$.  }

\db{
Note that our $ROI$-reconstruction approach can be applied {\it whenever the non-truncated cone beam  projection operator $D$ can be inverted} by an implementable formula or a ``blackbox algorithm" $Z$ that is well defined on smooth densities. Unlike other methods proposed in the literature we do not need any explicit restriction on the ROI location or any prior knowledge of the density in the ROI as long as an inverse $Z$ of the non-truncated cone beam projection $D$ exists in the sense stated above. As indicated above, the existence of such $\epsilon$-accurate inversion of the truncated cone beam projection $D_C$ in only guaranteed for $C$ relatively close to $B$.  
}

The iterative scheme~\eqref{def.algo} is formally similar to other iterative algorithms  also proposed in the literature for ROI reconstruction such as the so-called Iteration Reconstruction-Reprojection (IRR) algorithm~\cite{nassi:IRR82,kim:IRR85,Ziegler2008}, the Ordered Subsets Convex algorithm \cite{KamBee98} proposed to speed up CT reconstruction by reducing the number of projections and the iterative maximum likelihood (ML) algorithm proposed by Ziegler et al. \cite{Ziegler2008}
However, existing applications of the IRR method and other iterative methods for ROI reconstruction found in the literature are mostly heuristic and provide
no theoretical justification for convergence. In this paper, we provide a rigorous analysis of the inversion of the cone beam transform for sources located on a three-dimensional curve satisfying classical Tuy's condition. Using this theoretical framework, we prove that it is possible to define and compute an approximate inverse of the truncated cone beam transform. 

To validate our approach in the discrete setting, we have performed numerical experiments using four classical discrete x-ray acquisition geometries, with sources located on a sphere, a spiral, a circular curve and twin orthogonal circles. For each setting, we have simulated ROI-truncated cone beam  data acquisition using three different density functions in $\R^3$: a Shepp-Logan phantom, a mouse tissue density data sample, a human jaw density data sample. We have  performed extensive numerical tests using spherical ROIs with various centers and radii and found that the numerically computed `critical ROI radius' is relatively small as compared to the size of the support of $f$ and essentially insensitive to the ROI location.

\paragraph{Paper outline}
The paper is organized as follows. In Section~\ref{acquisition}, we recall the definitions of the ray and cone-beam transforms, and classical Tuy's condition
valid for acquisition settings with sources on smooth 3D curves. In Section~\ref{non.truncated}, we examine known inverse operators $Z$ implementing the reconstruction of densities from non-truncated projection data and study the continuity properties of $Z$ on adequate Sobolev spaces defined on the space of rays $\mathcal{R}_B$. In Section~\ref{regularization}, we define  a class of smoothing approximations of the identity in the image and projection domains, and we indicate how to implement these regularization operators \db{by `small' mollification}. In Section \ref{s.formal}, we describe our iterative ROI reconstruction algorithm from ROI-truncated data and we prove our main convergence results. 
In Section~\ref{sec.num}, we present numerical implementations of our iterative ROI reconstruction for discrete acquisition setups where sources are located on (1) a sphere, (2) a spiral, (3) a circular arm, (4) twin orthogonal circles, with simulated ROI-truncated x-ray data acquired from three densities in $\R^3$: a Shepp-Logan phantom, a mouse tissue density and a human jaw density. We analyze the accuracy of our ROI reconstruction approach and explore how the ROI radius impacts accuracy. Finally, we make some concluding remarks in Section~\ref{s.conclu}.

\section{X-ray projections and Tuy's condition} \label{acquisition}

We consider classical projection operators mapping density functions with domain in $\R^3$ into linear projections defined on appropriates spaces of rays. The most prominent examples of such projection operators are the ray transform and the cone-beam transform \cite{natterer:imagerec}.  
 
Recall that a {\it ray} $\tilde r(u,\theta)$ in $\R^3$ is a line passing through the point $u \in \R^3$ and parallel to the vector $\theta \in S^2$, where $S^2$ is the unit sphere of $\R^3$. That is $ \tilde r(u,\theta) = \{u + t \theta: t \in \R\}. $
A {\it half-ray} $r(a,\theta)$ in $\R^3$ is a half-line originating at the point $a \in \R^3$ and parallel to the vector $\theta \in S^2$. That is $r(a,\theta) = \{a + t \theta: t \ge 0\}. $

\subsection{The ray transform}   \label{s.raytra}

The {\it ray transform} maps a function $f \in L^1(\R^3)$ into its linear projections $X f$ obtained by integrating over rays at various locations and orientations, that is,
\begin{equation*}\label{xray}
 X f(u,\theta)=\displaystyle \int_{-\infty}^{\infty} f(u+t\theta) \,dt,
\end{equation*}
for $u \in \R^3$ and $\theta \in S^2$. Since $X f(u,\theta)$ does not change if $u$ is moved parallel to $\theta$,  it is sufficient to restrict $u$ to the plane through the origin that is orthogonal to $\theta$ in
$\mathbb{R}^3$, henceforth denoted by $T(\theta)$.
Thus, $X f$ is a function on the tangent bundle of the sphere that we
denote by 
$$\mathcal{T}=\{(u,\theta): \, \theta \in S^2, u \in T(\theta)\}.$$ 
Note that the pairs $(u,\theta)$ and $({u},-\theta)$
give the same ray $\tilde r(u,\theta)$, so that the mapping $(u,\theta) \to
\tilde r(u,\theta)$ is a double covering of $\mathcal T$ which can thus be viewed as
a 4-dimensional Riemannian quotient manifold. The
associated Riemannian volume element on $\mathcal{T}$ is $du \,
dQ(\theta)$, where $dQ(\theta)$ is the surface area on $S^2$ and
$du$ is the Lebesgue measure on the plane $T(\theta)$.

We will consider the action of the mapping $X$ on functions with compact support inside a fixed  open ball $B \subset \mathbb{R}^3$ of radius $\rho$ centred at the
origin. We denote by $\mathcal{T}_B$ the subset of $\mathcal{T}$ associated
with the rays passing through $B$, that is 
$$\mathcal{T}_B
=\{(u,\theta) \in \mathcal{T}: \tilde r(u,\theta) \cap B \ne
\emptyset\}.$$ 
Thus, $\mathcal{T}_B$ is an open submanifold of $\mathcal{T}$ with
compact closure in $\mathcal{T}$ 
and the natural
Riemannian volume element at $(u,\theta) \in \mathcal{T}_B$ is given by
$du \, dQ(\theta)$. We denote as $L^p(\mathcal{T}_B)$, $1 \le p \le \infty$, the
standard $L^p$ function spaces associated to this Riemannian volume.

\subsection{The cone-beam transform}
The {\it cone-beam transform} maps a function $f \in L^1(\R^3)$ into the function $D f$  defined by
\begin{equation*}\label{def.conebeam}
 D f(a,\theta)=\displaystyle \int_0^{\infty} f(a+t\theta) \,dt,
\end{equation*}
for $a \in \R^3$ and $\theta \in S^2$. Here, we view $a$ as the source of the half ray $r(a,\theta)$ with direction $\theta$. Hence, $D f$ is a function on the space of the half-rays
$$\mathcal{R}=\{(a,\theta): \, \theta \in S^2, a \in \R^3 \}.$$ 
The space $\mathcal{R}$ has the structure of a smooth  5-dimensional Riemannian manifold,  with  natural local coordinates defined by $a \in \R^3$ and  standard spherical coordinates on $S^2$. In particular $\mathcal{R}$ has a  Riemannian volume element  $d\mu = d a \,
dQ(\theta)$, where $dQ(\theta)$ is the surface area on $S^2$ and
$d a$ is the Lebesgue measure on $\R^3$.

We assume that all the unknown density functions $f$ have compact support inside 
a fix an open ball $B \subset \mathbb{R}^3$ of radius $\rho$ centered at the origin.
In the more realistic tomographic setups considered below, sources are located on a smooth bounded curve $\Gamma \subset \R^3$ supported outside the ball $B$. We denote by $\mathcal{R}_B$ the subset of all  half-rays with sources on $\Gamma$ and actually intersecting  $B$ that is
\begin{equation} \label{def.acrays} 
\mathcal{R}_B =\{(a,\theta) \in \mathcal{R}: r(a,\theta) \cap B  \ne \emptyset, \, \theta \in S^2, 
a \in \Gamma\}.
\end{equation}
We call $\mathcal{R}_B$ the set of {\it active rays}. $\mathcal{R}_B$ is a 3-dimensional manifold of class $C^{\infty}$ with natural local coordinates defined by the arclength parametrization $t$ of the curve $\Gamma$ and the standard spherical coordinates on $S^2$. Thus, $\mathcal{R}_B$ is a submanifold of $\mathcal{R}$ with
Riemannian volume element at $(a,\theta) \in \mathcal{R}_B$  given by 
$dt \, d\theta)$, where $d t$ is the Lebesgue measure on $\R$. The total volume $m(B) = \mu(\mathcal{R}_B)$ is clearly finite.

\paragraph{Sources on a curve: Tuy's condition.}

When the ray sources are located on a piecewise smooth curve $\Gamma$ exterior to a bounded open ball $B$, classical {\it Tuy's condition} on $\Gamma$ and $B$ (see \cite{Tuy:inversion,natterer:imagerec}) ensures that every  smooth function $f$ with compact support included in $B$ can be recovered from its non-truncated  cone-beam projection $Df$.

\begin{definition} \label{tuy}
Let $B$ be an open ball of finite radius centred at the origin.
 Let $\Gamma \in \R^3 \setminus B$ be a $C^{\infty}$ curve of length $L$ parametrized by  $t \to  \gamma(t) \in \R^3$, for $0 \leq t \leq L $, with non zero velocities $\gamma'(t)$. 

$\Gamma$ is said to verify  {\it strong Tuy's condition} if there is a $C^1$ function 
$\lambda = \lambda(x,\theta)$ defined for $(x, \theta) \in  \bar{B}\times S^2 $ and with values in $[0, L] $ such that,  for all $(x, \theta) \in  \bar{B}\times S^2$,
\begin{equation} \label{tuycondition}
\ip{ \theta}{ \gamma(\lambda(x,\theta) }  = 0 \quad  \text{and} \quad \ip{ \theta}{  \gamma'(\lambda(x,\theta)) }  \neq  0.
\end{equation}
Note that, by the Implicit Function Theorem, the function  $\lambda$ is  of class $C^{\infty}$.
\end{definition}

The strong Tuy condition is satisfied, for instance, when $\Gamma$ is a long enough circular helix "containing" the ball $B$, or when $\Gamma$ is the union of two  concentric circles positioned on orthogonal planes in $\R^3$.

\db{Even though an helix does not necessarily satisfy strong Tuy's condition
(in general, there are planes that intersect a helix at one point, with tangential intersection), a bounded circular helix is complete (in the sense of Tuy) as long as the support of the density is sufficiently small, and is surrounded by the helix. If this assumption holds, then tangential intersections between the planes and source curve are negligible as they occur on a set of Lebesgue measure zero \cite{KatIJMMS03}.
}

As mentioned above, we will consider in Section~\ref{sec.num} discrete applications of the cone-beam transform
for different practical acquisition setups including the {\it spherical case}, where $\Gamma$ is a sphere surrounding the target ball $B$, the {\it spiral case}, where  $\Gamma$ is  a segment of circular  helix, the {\it C-arm case}, where $\Gamma$ is  a circular arc and the {\it twin orthogonal circles case}, where $\Gamma$ is composed of two  concentric circles positioned on orthogonal planes in $\R^3$. In all these cases there is a formula to reconstruct a compactly supported smooth density function $f$ from its non-truncated projections.

\section{Reconstruction from non-truncated projections} \label{non.truncated}

For the non-truncated projection operators considered above, which map density functions in $\R^3$ into a full set of linear projections, it is possible in many classical cases to define a formal inverse operator.

For the non truncated ray transform $X$, when $f$ is in the space  $\mathcal{S}$ of functions on $\R^3$ having fast decreasing derivatives of all orders  and when $X f(u,\theta)$ is known for all  $(u,\theta) \in \mathcal{T}$, then there exists an inverse operator $Z$ such that $f(x) = Z . X f(x)$ (cf.~\cite[Sec. 2.2]{natterer:imagerec} or \cite{Helgason}).

For the non truncated cone-bean transform $D$, if the source location $\Gamma$ is a piecewise $C^{\infty}$ curve exterior to a ball $B$ and verifying strong Tuy's condition, then  for all $f $ in $C^2(\R^3)$ with compact support included in $B$,  there is an inverse operator $Z$ such that  $Z . D f(x)=f(x)$, and $Z$ can be implemented by one of several variants of Grangeat's formula  \cite{grangeat:1991,natterer:imagerec}. 
For example, in {\it spiral tomography}, where $\Gamma$ is a segment of a 
 circular  helix, the inverse $Z$ of the non truncated cone beam transform $D$ can be computed either by a variant of  Grangeat's formula or alternatively by the Katsevitch's formula \cite{Katsevich:spiralCT04, YuWang:spiral04}. In the setting of {\it C-arm tomography}, where $\Gamma$ is an arc of  circle, an approximate inverse operator $Z$ of the non truncated cone beam operator $D$ can be computed using again  a variant of Grangeat's formula \cite{Tuy:inversion, Zhao:unifiedconebeam}.

We point out  that all these {\it exact} formulas inverting the non truncated cone beam operator require smoothness conditions on $f$ to reconstruct $f$ as a {\it function}. As noted by Natterer \cite{natterer:imagerec}, Tuy \cite{Tuy:inversion} and other authors, to define the most generic linear operator $Z$ inverting the transform $f \to D f$, one should consider $f$ and $g = D f$ as distributions instead of functions. However, numerical reconstructions of $f$ from discretized projection data $ D f$ usually smooth the non truncated data $D f$ before reconstruction. Therefore classical proofs of exact inversion formulas for non truncated cone beam data tend to focus on {\it smooth} density functions $f$. Indeed, in spiral tomography, where $\Gamma$ is an helix, the original proof of Katsevich's inversion formula in \cite{Katsevich:spiralCT04} requires $f \in C^{\infty}_0(B)$; \db{finite degree of smoothness can be achieved using more sophisticated arguments \cite{katsevich2005stability}.} Similarly, when $\Gamma$ is a smooth curve, the proof of Grangeat's inversion formula in \cite{grangeat:1991} requires $f \in C^2(B)$. Also in the more academic setting of the ray transform, where the full set of projections (for all $(u,\theta) \in \mathcal{T}$) is known, the inversion formulas in \cite{natterer:imagerec,Helgason} require the Fourier transform of $f$ to decrease rapidly at infinity.

In the following, we will define exact inverses $Z$ of the non-truncated projection operators  $D$ 
as explicit linear operators acting on Sobolev spaces of densities. This definition will be useful to derive important {\it continuity} properties of $Z$.

We start by defining appropriate Banach spaces to handle the space of rays. 

\subsection{Banach spaces of smooth functions on manifolds} \label{smoothspaces}
Let $\mathcal{M}$ be a Riemannian manifold of class $C^{\infty}$ with volume element $d\mu$ and finite volume $\mu(\mathcal{M})$. Here we  consider only manifolds which are  either compact or are the interior of a compact manifold  with a $C^1$-boundary. One can then find and fix a {\it finite covering} of $\mathcal{M}$  by open relatively compact sets $U_j, j \in J$ endowed with diffeomorphic local maps $h_j: V_j \to U_j$, where the $V_j$ of are bounded open balls in  $\R^3$,  and each $h_j$ is the restriction to $V_j$ of a local map defined on an open Euclidean ball containing the closure of $V_j$. 
Explicit such finite coverings $U_j$  are easily specified for the manifolds of rays  $\mathcal{R}_B$ given by~\eqref{def.acrays}, and for $\Gamma \times S^2$, where $\Gamma$ is either a piecewise $C^{\infty}$ bounded curve in $\R^3$ or a whole sphere in $\R^3$.

Fix as above a finite covering $U_j, j \in J$, of $\mathcal{M}$ and the local maps 
$h_j: V_j \to U_j$. A function $g$ on $\mathcal{M}$  is said to be  uniformly bounded if and only is all the $g \circ hj$ are bounded. For any  $r > 0$, call $C^r(\mathcal{M})$  the space of all functions on $\mathcal{M}$ having continuous and uniformly bounded differentials  of all orders up to $r$.
For $1 \leq p \leq + \infty $, we denote $L^p(\mathcal{M})$ the usual Banach spaces of functions $g$ on $\mathcal{M}$ such that $|g|^p$ is $\mu$-integrable and $\mu$ is the Borel measure on $\mathcal{M}$.\\
For each $r$, the space $C^r (\mathcal{M})$ is included in the Sobolev space $W^r (\mathcal{M})$ of functions $g \in L^2(\mathcal{M})$ endowed with the Banach space norm 
\begin{equation*} \label{eq.norml}
\norm{g}_{W^{r}(\mathcal{M})} = \sum_{j \in J} \norm{g \circ h_j}_{W^r(V_j)} 
\end{equation*}

We have the following standard result (cf.~\cite{natterer:tomography,natterer:imagerec}).
\begin{proposition} \label{boundsX}
Let $\Gamma \subset \R^3$ be either a smooth bounded curve or a full sphere exterior to   a bounded open ball $B$. Let  $\mathcal{R}_B$ be the manifold of half-rays with sources on $\Gamma$ and intersecting $B$. For each integer 
$r \geq 0$, the non-truncated cone-beam transform $D$ is a bounded linear operator from $W^r(B)$ into $W^r(\mathcal{R}_B)$, as well as  from $L^{\infty}(B)$ into $L^{\infty}(\mathcal{R}_B)$, and maps $C^{r}(B)$ into $C^{r}(\mathcal{R}_B)$.  
The non-truncated ray transform $X$ is also a bounded linear operator from $W^r(B)$ into $W^r(\mathcal{T}_B)$ as well as from $L^{\infty}(B)$ into $L^{\infty}(\mathcal{T}_B)$, and maps $C^{r}(B)$ into  $C^{r}(\mathcal{T}_B)$.
\end{proposition}

\subsection{Inversion of the non-truncated ray transform }\label{sec.parallel.inverse}
An analytic inversion formula for the ray transform can be derived
from the classical Fourier slice theorem. In this section, we derive explicit
continuity properties for this inversion formula.

For any $\theta$ in $S^2$, let $T(\theta)$ be the plane orthogonal to $\theta$ in $\R^3$ and containing the origin. As seen in Sec.~\ref{s.raytra}, any ray $r(u,\theta)$ is non-ambiguously indexed by $\theta \in S^2$ and $u \in T(\theta)$ and the tangent bundle $\mathcal{T}$ of the unit sphere in $\R^3$ is a 4-dimensional manifold with  volume element $du \, d\theta)$.

Fix a an open ball $B$ of radius $\rho$ in $\R^3$ and let $\mathcal{T}_B$ be the manifold of all rays intersecting $B$. For any function $g(u, \theta)$ on $\mathcal{T}_B$, let $g_{\theta}$ be the function defined on $T(\theta)$ by $g_{\theta}(u) = g(u,\theta)$. For $v \in T_{\theta}$, the 2-dimensional Fourier transform of $g_{\theta}$ on the tangent plane $T(\theta)$ is given by
\begin{equation} \label{notation}
\F_{\theta} \, g_{\theta}(v) = \int_{T(\theta)} e^{- i \ip{u}{v}} \, g(u, \theta) \, du.
\end{equation}
whenever the integral is well-defined. 
Using standard inequalities, a direct computation shows that for  $g \in W^4(\mathcal{T}_B)$ then, for any $\theta \in S^2$ and $u \in T(\theta)$,
\begin{equation} \label{boundfourierG}
| \F_{\theta} \, g_{\theta} (v) | \leq 
c \, (1 + |v|^4)^{-1} \norm{ g}_{W^4(\mathcal{T}_B)},
\end{equation}
where the constant $c$ depends only on $\rho$ and not on $g$.

For $f \in L^2(B)$, the usual 3-dimensional Fourier transform of  $f$ will be denoted by
$$
\hat{f}(z) = \F f(z) = \int_{B} e^{- i \ip{z}{x}} \, f(x) \, dx, \quad \text{ for } z \in \R^3.
$$
By the Fourier slice theorem (cf. \cite[Sec. 2.2]{natterer:imagerec}), for any $\theta \in S^2$ and $z \in \R^3$ such that $\ip{z}{\theta}$= 0, the ray transform $g = Xf$ of  $f$ verifies 
\begin{equation} \label{Fourier.slice}
\hat{f}(z) = \F g_{\theta}(z), 
\end{equation}
{\it provided the two Fourier transforms involved in the formula are well defined}. As shown in \cite{natterer:imagerec,Helgason}  when  $\hat{f}(z)$ tends to zero at infinity faster than any polynomial in $z$, then equation \eqref{Fourier.slice} can be used to derive inversion formulas
to reconstruct $f$ from its non-truncated projections $g$.

For the non-truncated ray transform $X$, we now specify a {\it bounded} linear inverse defined on $W^4(\mathcal{T}_B)$. A function $g$ defined on $\mathcal{T}_B $ can be extended to $\mathcal{T}$ by setting $g=0$ on $\mathcal{T} \setminus \mathcal{T}_B $

\begin{proposition} \label{prop.inverse.parallel}
Fix a ball $B$ and define $\mathcal{T}_B $ as above. 
 Fix any Borel measurable function $z \to \theta(z)$ from $\R^3$ to $S^2$ such that $\ip{z}{ \theta(z)} =0$ for almost all $z \in \R^3$. For any $g \in W^4(\mathcal{T}_B)$ and all $x \in \R^3 $, the  following integral is necessarily finite:
\begin{equation} \label{parallel.inverse}
Jg (x) = (2 \pi)^{-3} \, \int_{\R^3} e^{i \ip{x}{z}} \, \F\, g_{\theta(z)} (z) \, dz.
\end{equation}
The restriction $Z g = 1_B Jg$ of $J g$ to the ball $B$ defines then a {\it bounded} linear operator $Z $ from $W^4(\mathcal{T}_B)$ into $L^{\infty}(B)$ and from $W^4(\mathcal{T}_B)$ into $L^2(B)$. Moreover, for any 
$f \in W^4(B)$, the non-truncated ray transform $g= Xf $ verifies the identity 
$f = Z X f $.
\end{proposition}

\proof   For any $g \in W^4(\mathcal{T}_B)$, the inequality~\eqref{boundfourierG} holds 
for all  $\theta  \in S^2$ and $v$ in the tangent plane $T(\theta)$. Hence for all $x \in \R^3 $, the integral $Jg(x)$, defined by
equation \eqref{parallel.inverse}, is bounded by  
\begin{eqnarray*}
|Jg (x) |  &\leq&  c \,  \norm{ g}_{W^4(\mathcal{T}_B)}  \, 
\int_{\R^3} \,  (1 + | z |^4)^{-1} dz \\
&  \leq& c \, \norm{ g}_{W^4(\mathcal{T}_B)},
\end{eqnarray*}
where the constant $c$ (changing from line to line) depends only on the radius of $B$. 

It follows that, for all  $g \in W^4(\mathcal{T}_B)$, there is a new constant $c$ such that the function  $Zg= 1_B \, Jg $ verifies 
\begin{equation} \label{boundXinverse }
\norm{ Z g }_{L^{\infty} (B)} \leq c \, \norm{ g}_{W^4(\mathcal{T}_B)}.
\end{equation}
Then $g \to Zg$ is a bounded linear operator from $W^4(\mathcal{T}_B)$ into $L^{\infty}(B)$ and hence also into $L^2(B)$. Moreover, for any $f \in W^4(B)$, the function $g = Xf$ is in $W^4(\mathcal{T}_B)$. Therefore the Fourier slice formula \eqref{Fourier.slice} combined with \eqref{parallel.inverse} show  that 
$$f = 1_B Jg = Zg= Z X f.$$ 
This achieves the proof. \qed

\subsection{Inversion of the non-truncated cone-beam transform}

We now construct an operator inverting the non truncated cone-beam transform $D$ when the projections belong to a Sobolev space of rays.

Fix a ball $B$ and define $\mathcal{T}_B $ as above. Let 
$\Gamma \subset \R^3$ be a $C^{\infty}$ curve with support exterior to  the open ball $B$ and parametrized by $ \gamma: \left[ 0, L \right] \to \R^3$. Assume strong Tuy's condition is verified, i.e., there is a $C^1$ function 
$\lambda = \lambda(x, \theta): \bar{B} \times S^2 \to [0, L] $ verifying  \eqref{tuycondition}.

Consider any function $g \in C^2(\mathcal{R}_B)$ with compact support inside $\mathcal{R}_B$. We extend $g$ to $\Gamma \times S^2$ by setting $g = 0$ on $\Gamma \times S^2 \setminus \mathcal{R}_B$ and then extend $g$ to a  $C^2$ function $G$ defined on $\Gamma \times \R^3$ by 
\begin{equation} \label{conebeam}
G(s,y)  = \norm{ y}^{-1} \,  g( s, \frac{y}{\norm{y} }), \quad \text{for all }   s \in  \Gamma , y \in \R^3 \setminus \{0\}.
\end{equation}
Now, for all $t \in [0, L]$, $y \in  \R^3 \setminus \{0\}$ we set
\begin{equation} \label{K}
K(t,y) = \frac{d}{dt} \partial_y G(\gamma(t),y)
\end{equation}
and, hence, for all $x \in B$, we  define the function $Z g$  by   
\begin{equation} \label{grangeatformula}
Z g (x) =  -\frac{1}{8 \pi^2}  \int_{\a \in T(\theta)} \int_{\theta \in S^2} 
 \frac{ \ip{ \theta, K(\lambda(x,\theta)}{\a) } }{( \ip{ \theta}{  \gamma(\lambda(x, \theta) ) }}  d \theta  d \a, 
\end{equation}
where, as above, $T(\theta) \subset S^2$ is the set of all $\a \in S^2$ such that  
$\ip{ \a}{ \theta } = 0$.

We have the following result.

\begin{proposition} \label{prop.inverse.curve}

The Grangeat formula~\eqref{grangeatformula} defines a linear operator 
$g \to Z g$ from $C^2(\mathcal{R}_B)$ into $L^{\infty}(B)$. Moreover there is a constant $c$ depending only on $\Gamma$ and the radius of $B$ such that, for all $g \in C^{2}(\mathcal{R}_B)$ with finite Sobolev norm 
$\norm{ g }_{W^4(\mathcal{R}_B)}$, we have 
\begin{equation} \label{granbound}
\norm{ Z g }_{L^{\infty}(B)} \leq c \norm{ g }_{W^4(\mathcal{R}_B)}.
\end{equation}
In particular $Z$ can be extended to a bounded linear operator from $W^4(\mathcal{R}_B)$ into $L^{\infty} (B)$ such that whenever $g = D f$ is the non truncated cone-beam transform of $f \in W^4(\mathcal{R}_B)$, one has the identity $ f = Z g = Z D f $. 
\end{proposition}

\proof When $g = Df$ with $f \in C^2(B)$, the assertion $Zg = f$ is proved with different notations in \cite[Sec.~5.5.2]{natterer:imagerec} using a variant of the  Grangeat's inversion formula due to Zeng, Clack and Gullberg   \cite{ZengClarkGullberg1994}.

For a generic $g$ in $C^{2}(\mathcal{R}_B)$, the vector valued function $K$, given by \eqref{K}, is continuous by construction  and hence remains bounded in $\R^3$ for $ t \in [0, L] $,  $y \in  S^2$. 

In the following, $ c, c_1, c_2, \ldots $ denote positive constants which depend {\it only} on the radius of $B$ and  $\Gamma$ but not on $g$. 

In equation \eqref{grangeatformula}, the denominator 
$\mbox{den}(x, \theta)  = \ip{ \theta}{  \gamma(\lambda(x, \theta) ) }$ is continuous for $x  \in \bar{B}$, $\theta \in S^2$ and is never zero due to Tuy's conditions, so that $| \mbox{den} | \geq c > 0 $ for some constant $c$. Then equation \eqref{grangeatformula} readily provides  a constant $c_1$ such that 
\begin{equation} \label{c1}
\norm{ Z g }_{ L^{\infty}(B)} \leq c_1 \sup_{t \in [0, L], \a \in  S^2} \norm{ K(t, \a)}_{\R^3}.
\end{equation}
Set $h(t, \theta) = g(\gamma(t), \theta)$. Equations \eqref {conebeam} and  \eqref{K} show that there is  a constant $c_2$ such that for all $g$ in $ C^2(\mathcal{R}_B)$,
\begin{equation} \label{c2}
\norm{ K(t, \a) }_{\R^3}  \leq c_2 \sup_{t \in [0,A], \theta \in  S^2} \norm{ \frac{d}{dt} \partial_{\theta} h (t, \theta) }_{\R^3},
\end{equation}
for all   $t \in [0,L], \a \in  S^2$.
By definition of $ W^{4}(\mathcal{R}_B)$, there is a constant $c_3 $ such that for any function  $g$ in $ W^{4}(\mathcal{R}_B)$ the function $h$ verifies 
\begin{equation} \label{c3}
\norm{ h }_{W^{4}([0, A] \times S^2)} \leq c_3 \, \norm{ g }_{ W^{4}(\mathcal{R}_B)}.
\end{equation}
The Sobolev imbedding theorem in dimension 3 holds on the Riemannian manifold $\mathcal{R}_B$, relating the norm of $h$ in $C^2(\overline{\mathcal R}_B)$
with its norm in $W^{4}(\mathcal{R}_B)$, as explained in Appendix~\ref{app.sobolevimbed}.  It thus provides a constant $c_4$  such that, for any function $h \in W^{4}([0, L] \times S^2)$, all partial differentials $ \frac{d}{dt} \partial_{\theta} h$ of order $\leq 2$ of $h$ are  bounded and continuous on $[0, L] \times S^2$  and verify 
\begin{equation} \label{c4}
\sup_{(t, \theta) \in [0,L] \times  S^2} \left|  \frac{d}{dt} \partial_{\theta} h(t, \theta)  \right|    \leq c_4 \, \norm{ h }_{W^{4}([0, L] \times S^2)}.
\end{equation}
Combining the inequalities \eqref{c1} \eqref{c2} \eqref{c3} \eqref{c4}, we get
\begin{equation*} 
\norm{ Z g}_{ L^{\infty}(B)} \leq c_1 c_2 c_3 c_4  \norm{ g }_{ W^{4}(\mathcal{R}_B)}
\end{equation*}
which achieves the proof. \qed
\paragraph{Remark: Katsevich's inversion formula.} As mentioned above, in spiral tomography, the source curve
$\Gamma \subset \R^3$ is a circular helix, which can be parametrized as $\gamma(t) = ( a\cos(t), a\sin(t) , k t )$ for some fixed  positive $a, k$. Let $f$ be a compactly supported density 
function $f \in C^{\infty}_0 (B)$, with $\rho < a$ (so that the helix is surrounding the support of $f$). Katsevich proved (cf. \cite{Katsevich:spiralCT02})  that, in this setting, $f$ can be reconstructed from its non-truncated cone-beam projections $D f$ by a formula which can be written as 
$$f = \frac{1}{2}(U + V). D f,  $$ 
where $U$ and $V$ are explicit  operators involving {\it divergent}  integrals. This divergence is carefully analyzed  by adequate approximations in   \cite{Katsevich:spiralCT02}, but this inversion formula  remains rather unwieldy and the continuity properties of $(U+V) $ are not easy to evaluate  directly. 

\db{As mentioned above, even though an helix does not  satisfy strong Tuy's condition in general, this condition holds for a bounded circular helix as long as the support of the density
is sufficiently small.} Hence, we can define an explicit inverse operator $Z$ of the non-truncated operator $D$ according to Proposition~\ref{prop.inverse.curve}. With respect to  Katsevich's inversion formula, our approach has the advantage of reconstructing $f$ from projection data $D f \in W^{4}(\mathcal{R}_B)$ for  all $f \in W^4(B)$ and to provide a precise Sobolev continuity property for the inverse operator $Z$.

\section{Regularization in the space of rays }\label{regularization}

We now define and construct a class of regularization operators in the space $L^2(\mathcal{R}_B)$. We start by defining a notion of {\it approximations of the identity} on the manifolds considered in Section~\ref{non.truncated}.

\begin{definition} \label{approxID} 
Let $\mathcal{M}$ be a $C^\infty$ Riemannian manifold with volume element $d\mu$ and finite volume. As above, assume that $\mathcal{M}$ is either compact or is the interior of a compact manifold with a $C^1$-boundary. For any integer $r \geq 1$, we call  {\it $C^{r}$ approximation of the identity} in $L^2(\mathcal{M}, \mu)$ any sequence of linear operators 
$\tau_N: L^2(\mathcal{M}) \to C^{r}(\mathcal{M})$ verifying the following conditions. 
\begin{itemize}
\item[(i)] There is a constant $c$ such that for all $g \in L^2(\mathcal{M})$ and all integers $N$ 
\begin{equation} \label{bound.tau.2}
\norm{\tau_N g }_ {W^r(\mathcal{M})} \leq c N^r \, \norm{ g }_{L^2(\mathcal{M})}.
\end{equation}
\item[(ii)] For any $g \in L^2(\mathcal{M})$ 
\begin{equation*} \label{limitL2}
\lim_{N \to \infty} \norm{ g -  \tau_N g }_{ L^2(\mathcal{M}) }= 0, \quad \text{for each }   g  \in L^2(\mathcal{M}).  
\end{equation*}
\item[(iii)] For each integer $2 \leq p \leq (r+1) $ there is a constant $c$ such that for all $ g  \in W^p(\mathcal{M})$
\begin{equation*} \label{limitWp}
\norm{ g -  \tau_N g }_{ W^{p-1} (\mathcal{M}) } \leq \frac{c}{N} \norm{ g }_{W^p(\mathcal{M})}. 
\end{equation*}
\item[(iv)]
Whenever $g$ has compact support then $\tau_N g$ also has compact support.
\end{itemize}

\end{definition}
Note that $N \to \tau_N $  will remain a $C^{r}$ approximation of the identity  in  $L^2(\mathcal{M}, \nu)$ for any positive Borel measure $\nu$ on $\mathcal{M}$ such that both densities $ \frac{d\nu}{d\mu}$ and $\frac{d\mu}{d\nu}$ are bounded.

We next show how to construct  $C^{r}$ approximations of the identity in  $ L^2(\mathcal{M}, \mu)$.

\subsection{Approximation of the identity by  small  convolutions}  When the manifold $\mathcal{M}$ is a bounded open Euclidean ball in $\R^k$, one can generate a  $C^r$ approximation of the identity as follows. Select any  fixed $C^r$ function $w \geq 0 $  on $\R^k $ with  compact support and Lebesgue integral equal to 1 and, for  $f \in L^2(\R^k)$, define the \textit{``small" convolutions} by 
$$\sigma_N f =  f *w_N, $$
where  $w_N(x) = N^k  w(N x)$. Standard results on convolutions show that the sequence $(\sigma_N)$ verifies  the properties (i),(ii) and (iv) of Definition~\ref{approxID}. The proof of property (iii) is the following.

For $z \in \R^k$ and $1 \le p < \infty$, we define $\phi_p(z) =  (1+ |z|^2)^{p/2}$. Denoting
by $W_N$ the Fourier transform of $w_N$, we have that $W_N(z) =\hat{w}(z/N)$, where $\hat{w}$ is the Fourier transform of $w$. Hence, for all $z \in \R^k$,
\begin{equation}  \label{ineq.w}
|W_N(z) -1| = |\hat w(z/N) -1 |\leq  \frac{c}{N} |z|, 
\end{equation}
where $c= \norm{(\hat w)'}_{L^{\infty}}$. From inequality~\eqref{ineq.w}, since $(\sigma_N f)^{\widehat{}} = W_N \hat f$, it follows that
\begin{eqnarray*}
 \phi_{p-1}(z) \, | (\sigma_N f)^{\widehat{}}(z) - \hat{f}(z) |  & = &
\phi_{p-1}(z) \, | W_N(z) \hat f(z) - \hat{f}(z) | \\
 &\leq& 
\frac{c}{N}  |z| \, \phi_{p-1}(z) \, |\hat{f}(z) | \\
&\leq& \frac{c}{N}  \phi_{p}(z)  \,  | \hat{f}(z) |,
\end{eqnarray*}
for all $z \in \R^k$. From the last inequality and by the definition of Sobolev norms, we then get:
\begin{eqnarray*}
\norm{ \sigma_N f - f }_{W^{p-1}(\R^k)} &=& \norm{  \phi_{p-1} \, ( (\sigma_N f)^{\widehat{}} -\hat{f}) }_{L^2(\R^k)} \\
&\leq& \frac{c}{N} \norm{ \phi_{p} \, \hat{f} }_{L^2(\R^k)} \\
& = & \frac{c}{N} \norm{ f }_{W^{p}(\R^k)}.
\end{eqnarray*}
This proves property (iii). \qed

Small convolutions are easily  and explicitly extended  to the manifolds of active rays $\mathcal{R}_B$, as we now show by patching together local small convolutions through appropriate local maps.

\begin{proposition}  \label{prop.app}
Let $\mathcal{M}$ be a $C^{\infty}$ Riemannian manifold of finite volume. As above, assume that $\mathcal M$ is either compact or is the interior of a compact manifold with  $C^1$-boundary. 
Then, for any integer $r$, one can construct explicitly  a $C^r$ approximation  of the identity $N \to \tau_N$ on $L^2(\mathcal{M})$.  
\end{proposition}

\proof  

As observed above, on $\mathcal{M}$ we can select an open finite covering $U_j, j \in J$, and local maps $h_j: V_j \to U_j$ to construct a finite partition of unity by $C^{\infty}$ functions $u_j$ with compact supports included in $U_j$ and verifying $0 \leq u_j \leq 1$ and $\sum_{j \in J} u_j = 1$. On each Euclidean ball $V_j$, select a $C^{r}$ approximation  of the identity  $N \to \sigma_N (j)$ in $L^2(V_j)$, for instance by small convolutions as indicated above.  

For any  $g  \in L^2(\mathcal{M})$, let  $g_j= g \, u_j$ and define the operators  $ g \to \tau_N g$ by
\begin{equation} \label{formula.tau}
\tau_N g =   \sum_j G_j \circ h_j^{-1},  
\end{equation}
where $G_j = \sigma_N (j)  (g_j \circ h_j)$
Since each map $h_j$ can be smoothly extended to a neighborhood of $\bar{V_j}$,  each  $h_j$ has bounded derivatives  of any order. Hence the mapping $g_j \to g_j \circ h_j$  is a bounded linear operator  from $L^2(U_j, \mu)$ to $L^2(V_j)$, and from $W^r(U_j)$ to $W^r(V_j)$. Similar boundedness properties hold for the linear operators mapping $G_j \to G_j \circ h_j^{-1}$ and $g \to g u_j$. The explicit formula \eqref{formula.tau} and the fact that the $\sigma_N(j)$ are $C^r$ approximations of the identity in $L^2(V_j)$ then implies directly  that the sequence $(\tau_N)$ satisfies the properties (i)-(iv) of Definition~\ref{approxID}. Hence $(\tau_N)$ is a  $C^{r}$ approximation  of the identity in   $L^2(\mathcal{M}, \mu)$. \qed

Proposition~\ref{prop.app} applies in particular to manifolds of active rays $\mathcal{R}_B$ associated  to an open ball $B$ and a smooth set of sources $\Gamma$ exterior to $B$.

\section{Reconstruction from ROI-truncated projections} \label{s.formal}

\begin{figure}[t]
\centering
\includegraphics[height= 3.5 in]{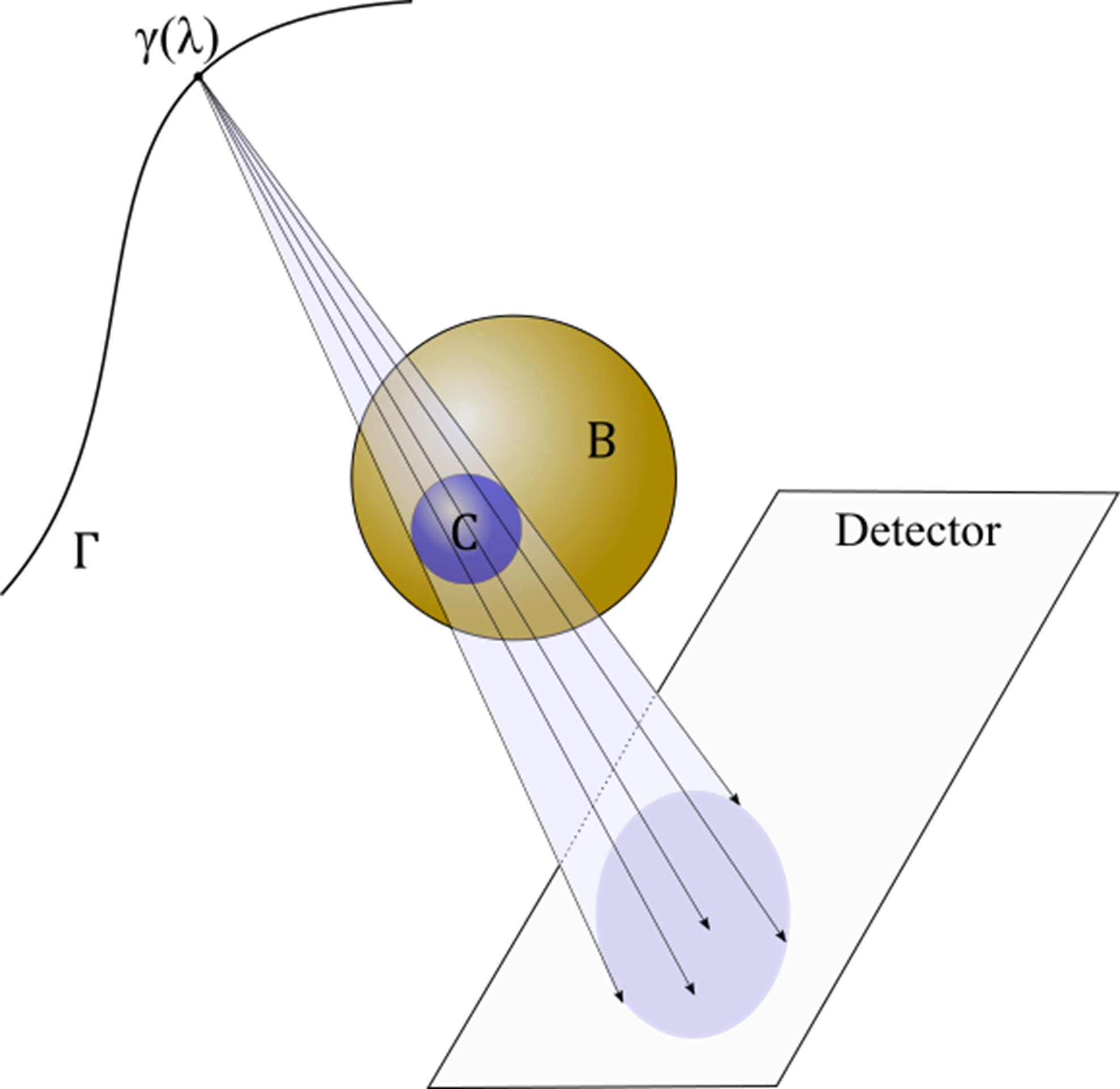}
\caption{ROI-truncated cone-beam acquisition: projections are restricted to half-rays intersecting the ROI, which is a ball $C$ included in the target ball $B$.}
\label{coll_xray}
\end{figure}

Fix  a bounded open ball $B \subset \R^3$ and a smooth set of sources $\Gamma$ verifying strong Tuy's condition. Let $C$ be  a spherical ROI strictly included in $B$. As illustrated in Figure~\ref{coll_xray}, the {\it $C$-truncated} cone-beam transform $D_C f$ of  $f$ is the restriction of $Df$ to the manifold $\mathcal{R}_C \subset \mathcal{R}_B $ of half-rays which intersect $C$. Denoting the indicator function of a set $G$ by $1_G$, one can then write  
\begin{equation} \label{def.D_C}
D_C = 1_{\mathcal{R}_C} D, \quad 
Y_C = 1_{ \mathcal{R}_B - \mathcal{R}_C} D \quad \text{and} \quad D = D_C + Y_C.
\end{equation}
by Proposition~\ref{boundsX}, $D_C$ is a bounded linear operator from $L^{\infty}(C)$ into $L^{\infty}(\mathcal{R}_C)$.

As observed above,  the non-truncated operator $D$ can be inverted by  a  linear operator $Z$ such that   $f = Z D f$  for all densities $f \in W^4(B)$ having compact support. However the $C$-truncated operator  $D_C$ {\it cannot be directly inverted} by applying $Z$ to $D_C f$, even within the region $C$.  So we now formally  define `approximate' inverses $Z_C$ for $D_C$ .

\db{
\subsection{Approximate inverses of ROI-truncated cone beam transforms}
\begin{definition} \label{eps.inverse}
Fix  a bounded open ball $B \subset \R^3$ and a smooth set of sources $\Gamma$ verifying strong Tuy's condition. Let $C$ be  a spherical ROI strictly included in $B$. For any  $\epsilon > 0$, we say that the $ROI$-truncated cone-beam transform $D_C$ admits an {\it $\epsilon$-accurate inverse}  $Z_C$ if $Z_C$ is a {\it bounded} linear operator from $L^{\infty}(\mathcal{R}_C)$  to $L^{\infty}(B) $ verifying 
\begin{equation*} \label{ZC}
\norm{(I - Z_C D_C) f }_{L^{\infty}(B)} \leq \epsilon \norm{ f }_{ W^5(B) }
\end{equation*}
for all $f \in W^5(B) $. 
\end{definition}
Note that the $\epsilon$-accurate inverse $Z_C$ of $D_C$ is in general {\it not unique}. 
Indeed recall that there are  constants $c$ and $c_1$ such that  for any  $f \in W^5(B)$ one has $\norm{f}_{L^{\infty}(B)} \leq c \norm{f}_{W^5(B)} $ and hence 
$$\norm{D_C f}_{L^{\infty}(\mathcal{R}_B)} \leq  c_1 \, \norm{f}_{L^{\infty}(B)} \leq   c_1 c \norm(f)_{W^5(B)}. $$
Fix now  any bounded linear mapping $K$ from $L^{\infty}(\mathcal{R}_B)$ into $L^{\infty}( B)$ with operator norm $\norm{K} < \frac{\epsilon}{c_1 c}$. For any $f \in W^5(B)$ we then have  $$\norm{ K D_C f}_{L^{\infty}(B)} \leq \epsilon \norm{f}_{W^5(B)}.$$ For any $\epsilon$-accurate inverse $Z_C$  of $D_C$ the operator $Z_C + K$ will then be a $2 \epsilon$-accurate inverse of $D_C$.
}

\subsection{Reconstruction from ROI-truncated projection data}

\db{
Let $B \subset \R^3$ be a spherical region and $\Gamma$ inside $B$ a smooth curve
verifying strong Tuy's condition. As seen above, the non truncated  cone-beam transform $D$ with sources on $\Gamma$ has then an inverse $Z : W^4(\mathcal{R}_B) \to L^{\infty}(B)$ such that  $Z D f = f$ for all $f$ in $W^4(B)$. The theorem below shows how to obtain an $\epsilon$ accurate inverse of the ROI truncated operator $D_C$.
\begin{theorem} \label{ROIalgo} ($ROI$-reconstruction algorithm). 
With the notation above,  let  $N \to  \tau_N$ be   any $C^4$-approximation of the identity in $L^2(\mathcal{R}_B)$ as in  Definition \ref{approxID}. For any sphere $C \subset B$ define the operators 
$$U_N =  Z \tau_N (D -D_C). $$
Given  any  $\epsilon > 0$, one can then find $N= N(\epsilon)$ and $\eta(\epsilon ) > 0 $ such that the operator $ U = U_{N (\epsilon)}$  becomes  a {\it contraction} of $L^{\infty}(B)$ provided  $\rad(B) - \rad(C) \leq \eta(\epsilon)$. \newline
With $N$ fixed as above, for any  
$g \in L^{\infty}(\mathcal{R}_B )$, define the functions  $f_j$ $j= 1, 2, \ldots$ by the recurrence 
\begin{equation}\label{d:algo}
f_{j+1} = f_0 + U f_j  
\end{equation}
with $f_0 = Z \tau_N g$. Then as $j \to \infty$ the sequence $(f_j)$ converge at exponential speed in $L^{\infty}(B)$ to a limit $Z_C g$. This defines a bounded linear operator  $Z_C$ from $L^{\infty}(\mathcal{R}_B )$ into $L^{\infty}(B)$. 
Moreover for  any unknown density $f$ in $W^5(B)$, one can use the $ROI$-truncated data $g= D_C f$ to compute an approximation of $f$ given by $\hat{f} = Z_C D_C f$ and verifying 
\begin{equation} \label{ZCDC}
\norm{f - \hat{f }}_{ L^{\infty}(B)} \leq \epsilon \norm{f}_{ W^5(B)}
\end{equation}
Hence $Z_C$ is an $\epsilon$ accurate inverse of the ROI truncated operator $D_C$.
\end{theorem}
}

\paragraph{Remark.} According to Theorem~\ref{ROIalgo}, the truncation region $C$ is strictly included  in $ B$  and must be large enough for the theorem to hold. The proof of theorem (presented below) provides an explicit lower bound $\rad(B) - \rad(C) \eta(\epsilon)$ implying  the existence of an $\epsilon$-accurate inverse for the $C$-truncated cone beam projector $D_C$. Our theoretical  estimate of $\eta(\epsilon)$ is clearly  too pessimistic. Our numerical tests (see  Sec.~\ref{sec.num}) indicate for instance that, for $\epsilon = 0.10$ and for all spheres  $C \subset B$ with same center as $B$ and radius $\rad(C) > rad(B) / 2$,  our $ROI$-reconstruction algorithm \eqref{d:algo} converges to an $\epsilon$-accurate inverse $Z_C$ of $D_C$. This is a favorable situation for radiation exposure reduction by $ROI$ truncated data acquisition combined with our $ROI$ reconstruction algorithm.

As seen above, $\epsilon$-accurate inverses are generally not unique. The following corollary outlines alternative constructions of  $Z_C$.

\begin{corollary} \label{variantalgo} The notations and hypotheses are the same as Theorem \ref{ROIalgo}. Fix two $C^4$-approximations of the identity: $(\tau_n)$ in $L^2(\mathcal{R}_B)$ and   $(\sigma_n)$ in    $L^2(B)$. For any $n \in \N$ and any spherical region $C \subset B$,  define the operator $\widetilde U_n = \sigma_n Z \tau_n (D-D_C)$. Then, given $\epsilon >0$, one can find $N(\epsilon)$ and $\eta(\epsilon) >0$ such that provided $rad(B) - rad(C) \leq \eta(\epsilon)$, the operator  $\widetilde U_{N(\epsilon)}$ is a contraction of $L^{\infty}(B)$.
Then for each $g \in L^{\infty}(B)$, the sequence $(f_j)$, given by the recurrence \eqref{d:algo} with $U =\widetilde U_{N(\epsilon)}$ and $f_0= \sigma_N Z \tau_N$, converges in $ L^{\infty}(B)$ to a limit $Z_Cg$, 
where $Z_C$ is  an $\epsilon$-accurate  inverse of the ROI-truncated transform $D_C$.
\end{corollary}

\subsubsection{Proof of the Theorem~\ref{ROIalgo}}

Before presenting the proofs, we need the following two lemmata. For the remaining of this section, let $\Gamma, B, C, D, D_C$ be given as above.

\begin{lemma}\label{normYC}
 There is  a constant $c$ determined by $\Gamma$ and $B$ only such that, for any  sphere  $C \subset B$ and any $f \in L^{\infty}(B)$, the linear operator $Y_C = D - D_C $ verifies
\begin{equation} \label{YC}
\norm{ Y_C f  }_{L^2(\mathcal{R}_B)} \leq c \left( \rad(B) - \rad(C) \right)^{1/2} \norm{  f  }_{L^{\infty}(B)},
\end{equation}
where $\rad(C)$ is the radius of $C$.
\end{lemma}
Proof:
Let $s$ be any source position on the curve $\Gamma$. Denote by $z(C)$ the center of $C$. Call $H(s,C) \subset S^2$ the set of all $\theta \in S^2$ such that the half-ray $r(s,\theta)$ intersects $C$. The set of all these half-rays is a cone of revolution  with vertex $s$,  axis $\left[ s, z(C) \right]$, and  half-aperture angle 
$0 < \alpha(s,C) <  \pi/2$. The area of the spherical cap $H(s, C)$ is  hence classically given by 
\begin{equation} \label{area}
\area( H(s,C) ) = 2 \pi (1 -\cos(\alpha(s,C)  ).
\end{equation}
Elementary geometry yields 
\begin{equation} \label{cos}
\cos( \alpha(s,C) ) = k( \rad(C) , | s- z(C) |),
\end{equation}
where   $k(u,v) = \frac{v}{ ( u^2 + v^2 )^{1/2}}\,$.

Since $s \in \Gamma$ and  $C\subset B$, the numbers $u = \rad(C)$ and  $v = |s - z(C) |$  remain respectively in the bounded intervals  $ [ 0,  \rad(B) )$ and $[ m, M ],$  where  
$$
0 < m = - \rad(B) + \min_{s \in \Gamma}  |s - z(B) | \, \text{ and } \,
M = \rad(B) +\max_{s \in \Gamma}  |s - z(B) |.  
$$
The function $(u,v) \to k(u,v)$ is $C^{\infty}$ on the rectangle 
$ J=[ 0,  \rad(B) ] \times [ m, M ]$. Hence there is a Lipschitz constant $c$ such that, for all 
$(u_1, v_1)$ and $(u_2, v_2)$ in $J$, one has
$$
| k(u_1,v_1) - k(u_2,v_2) | \leq c ( | u_1 - u_2 | + | v_1 - v_2| ).
$$ 
Due to equation~\eqref{cos}, this implies 
$$
|\cos( \alpha(s,C) ) - \cos( \alpha(s,B) ) | \leq c ( \rad(B) - \rad(C)) + |  ( | s - z(C) |  - | s - z(B)| ) |.
$$ 
Since  $ |  ( | s - z(C) |  - | s - z(B)| ) | \leq  | z(C) - z(B)| \leq \rad(B) - \rad(C) $, it follows that 
\begin{equation} \label{cosrad}
|\cos( \alpha(s,C) ) - \cos( \alpha(s,B) ) | \leq 2 c \, ( \rad(B) - \rad(C)).
\end{equation}
 The  total volume of $\mathcal{R}_C$ is
$$
\vol(\mathcal{R}_C) = \int_{\Gamma} \area( H(s,C) ) \,  ds, 
$$
where $d s d \theta$ is the volume element in the  manifold  of active rays $\mathcal{R}_C$.
Hence, due to \eqref{area}, we have
\begin{eqnarray*}
\vol(\mathcal{R}_B - \mathcal{R}_C)&=& \int_{\Gamma} \left(\area( H(s,B) - \area(H(s,C) \right) \, ds \\
&=&
2 \pi \, \int_{\Gamma}  (\cos (\alpha(s,C)  ) -\cos (\alpha(s,B)  ) \, ds. 
\end{eqnarray*}
From the last equation, using~\eqref{cosrad}, we obtain
\begin{equation} \label{RBRC}
\vol(\mathcal{R}_B - \mathcal{R}_C)  \leq 4 \pi  L  c \, ( \rad(B) - \rad(C)),
\end{equation}
where $L$ is the length of $\Gamma$.
The $L^2$ norm of the indicator function $1_{\mathcal{R}_B - \mathcal{R}_C}$ then verifies
$$
\norm{1_{\mathcal{R}_B - \mathcal{R}_C}}_{L^2(\mathcal{R}_B)}^2 = \vol(\mathcal{R}_B - \mathcal{R}_C)  \leq 4 \pi L c ( \rad(B) - \rad(C)). 
$$ 
Due to Proposition~\ref{boundsX}, there is a constant $c_0$ such that, for any $f \in L^{\infty}(B)$, one has  
$\norm{D f}_{L^{\infty} (\mathcal{R}_B)} \leq c_0  \norm{f}_{L^{\infty}(B)}$. Hence, since $C\subset B$ we have 
\begin{eqnarray*}
\norm{Y_C f}_{L^{2}(\mathcal{R}_B)} &=& \norm{1_{\mathcal{R}_B - \mathcal{R}_C} D f)}_{L^{2} (\mathcal{R}_B)} \\
&\leq&  
\norm{1_{\mathcal{R}_B - \mathcal{R}_C}}_{L^2(\mathcal{R}_B)} \norm{ D f}_{L^{\infty} (B)} \\
&\leq& 
c_1 ( \rad(B) - \rad(C))^{1/2}  \norm{f}_{L^{\infty} (B)},
\end{eqnarray*}
where $c_1 = c_0 (4 \pi L c )^{1/2}$. This achieves the proof when $\Gamma$ is a curve of length $L$ and  does not intersect the closure of $B$.  \qed

As seen above, the non-truncated cone beam transform $D$ can be inverted through a bounded linear operator $Z : W^4(\mathcal{R}_B) \to L^{\infty}(B)$ given by the Grangeat's formula \eqref{grangeatformula}. The lemma below shows how to construct a contraction on $L^{\infty}(B)$ from $Z$ and $D-D_C$.

\begin{lemma}\label{contraction}
Let   $(\tau_N)$ be a $C^4$ approximation of the identity in $L^2(\mathcal{R}_B)$ as in Definition~\ref{approxID}.  Then there is a constant $k$ such that for  any   $C \subset B$ with radius verifying  $\rad(B) - \rad(C) < k/N^8$, the linear operator  $U_N = Z \tau_N Y_C = Z \tau_N  (D - D_C)$  is a  {\it contraction} from $L^{\infty}(B)$ into $L^{\infty}(B)$,  with operator norm  $\norm{ U_N }_{L^{\infty}(B)} \leq 0.9$. 
\end{lemma}

\proof
By inequality \eqref{YC}, there is a constant $c$ such that, for all $C \subset B$ and all $ f  \in L^{\infty}(B)$, the operator $Y_C = D-D_C$ verifies
$$
\norm{ Y_C f  }_{L^2(\mathcal{R}_B)} \leq c  \, (\rad(B) - \rad(C) )^{1/2} \norm{  f  }_{L^{\infty}(B)}. 
$$
Applying inequality \eqref{bound.tau.2} to $g = Y_C f$ with $r=4$, we obtain  a new constant $c_1$ such that, for all  $C \subset B$, all $ f  \in L^{\infty}(B)$ and all $N$
\begin{eqnarray*}
\norm{ \tau_N Y_C f  }_ {W^4(\mathcal{R}_B)} & \leq&  c_1  N^4 \norm{ Y_C f  }_{L^2(\mathcal{R}_B)} \\
&\leq&  c_1 c N^4  \, ( \rad(B) - \rad(C) )^{1/2} \, \norm{  f  }_{L^{\infty}(B)}. 
\end{eqnarray*}
By applying inequality \eqref{granbound} to the function $\tau_N Y_C f$, we then obtain  a new constant $c_2$ such that, for all  $C \subset B$, all $ f  \in L^{\infty}(B)$ and all $N$,
\begin{eqnarray*}
\norm{ U_N f }_{ L^{{\infty}} (B)} &=&  \norm{ Z \tau_N Y_C f  }_{ L^{\infty} (B)} \\
&\leq & 
c_2 \norm{ \tau_N Y_C f  }_ {W^4(\mathcal{R}_B)} \\
&\leq&
c_2 c_1 c  N^4   \, ( \rad(B) - \rad(C) )^{1/2} \norm{f }_{L^{\infty}(B)}.
\end{eqnarray*}
Set $c_3= c_2 c_1 c $. Then, provided $ \rad(B) - \rad(C)  \leq c_3/N^8$, the linear operator $U_N$  is a contraction of $L^{\infty} (B)$ with operator norm $\norm{ U_N }_{L^{\infty}(B)} \leq 0.9$.
\qed \\

We can now prove Theorem~\ref{ROIalgo}.

Select and fix a $C^4$ approximation of the identity $N \to \tau_N)$ in $L^2(\mathcal{R}_B)$, so that the bounded operators $\tau_N : L^2(\mathcal{R}_B) \to W^4(\mathcal{R}_B)$ verify definition \ref{approxID}. We will use the following shorthand notations for the  norms of various linear operators $T$: 
\begin{eqnarray*}
& |T|_{W5W5}  & \text{ is the norm of } T: W^5(B)  \to W^5(\mathcal{R}_B) \\
& |T|_{W5W4}  & \text { is the norm of }  T: W^5(\mathcal{R}_B)  \to W^4(\mathcal{R}_B)   \\ 
& |T|_{W4L\infty}  &\text{ is the norm of }  T: W^4(\mathcal{R}_B) \to L^{\infty}(B) \\
& |T|_{L2W4}    &\text{ is the norm of }  T: L^2(\mathcal{R}_B) \to W^4(\mathcal{R}_B)   
\end{eqnarray*}

Due to Propositions \ref{boundsX} and \ref{prop.inverse.curve}, Definition \ref{approxID}  and equation \eqref{grangeatformula}, there is a constant $c >0$ such that for all integers $N$
\begin{equation} \label{Z1}
| D |_{W5W5} < c, \;  | Z |_{W4L\infty}   < c,  \;  | I - \tau_N |_{W5W4}\leq c/N,  \;  
| \tau_N |_{L2W^4} \leq c N^4. 
\end{equation}
Given an $ \epsilon >0 $, fix an integer $N = N(\epsilon)$ by
 \begin{equation} \label{Neps}
N = N(\epsilon) \equiv  10 c^3 / \epsilon.
\end{equation}
Since $N = N(\epsilon)$ is now fixed, we will write $\tau= \tau_N$. By Lemma \ref{contraction}, there is a constant $k$ such that for any  spherical region  $C \subset B$ verifying $\rad(B) - \rad(C)  \leq  k / N(\epsilon)^8 $, the operator  $U_N=U = Z \tau Y_C$  is a  contraction from $L^{\infty}(B)$ into $L^{\infty}(B)$,  with operator norm $\norm{ U }_{L^{\infty}(B)} <  0.9$. 

We now fix  $\eta(\epsilon) = k / N(\epsilon)^8 $ and assume that the ROI radius  verifies $\rad(B) - \rad(C) \leq \eta(\epsilon)$, which  forces $U$ to be a contraction  with norm inferior to $0.9$.
Given any $g$ in $L^{\infty}(\mathcal{R}_B)$, define a sequence $(f_j) \subset L^{\infty}(B)$ by the algorithm \eqref{d:algo}. By construction one has then 
$$
\norm{ f_{j+1} - f_j }_{L^{\infty}(B)}\leq \norm{U} \norm{ f_{j} - f_{j-1} }_{L^{\infty}(B)} \leq 
0.9 \norm{ f_{j} - f_{j-1} }_{L^{\infty}(B)}.
$$
Hence,   as $j \to \infty$, the sequence $(f_j)$  converges  at exponential speed in $L^{\infty}(B)$ to a limit $Z_C g \in L^{\infty}(B)$. By \eqref{d:algo}, we must have 
\begin{equation} \label{A1}
Z_C g = f_0 +U Z_C f = Z \tau g + U Z_C g
\end{equation}
This defines a linear operator $Z_C: L^{\infty}(\mathcal{R}_B) \to L^{\infty}(B)$. We now show that $Z_C$ has bounded operator norm.  Since $\norm{ U }_{ L^{\infty}(B) } \leq 0.9$, the operator $(I - U): L^{\infty}(B) \to L^{\infty}(B) $ has a bounded inverse given by the converging series $ \sum_{j = 0}^{\infty} U^j$, which yields
$$
\norm{ (I-U)^{-1}}_{ L^{\infty}(B) } \leq \frac1{1 - 0.9} =10.
$$
Equation   \eqref{A1} yields that, for all $g \in L^{\infty}(\mathcal{R}_B)$,
\begin{equation} \label{A2}
Z_C g=   (I-U)^{-1} Z \tau g.
\end{equation}
Hence, due to the bounds \eqref{Z1},   
\begin{eqnarray} \label{normZC}
\norm{Z_C g }_{ L^{\infty}(B) } &\leq&  \norm{ (I-U)^{-1} }_{ L^{\infty}(B) } | Z |_{W4L\infty}
 | \tau |_{L2W4} \norm{g}_{L^2(\mathcal{R}_B)} \nonumber \\
 &\leq & 
10 c^2 N(\epsilon) ^4 \norm{g}_{L^2(\mathcal{R}_B)} \nonumber \\
&\leq & 
10 c^2 N(\epsilon) ^4 m(B)^{1/2} \norm{g}_{L^{\infty}(B)}
\end{eqnarray}
where $m(B)$ is the finite Riemannian volume of $\mathcal{R}_B$. So the operator norm of $Z_C$ is bounded.

For $f \in W^5(B)$ we have $f = Z D f $ and the $ROI$ truncated transform $h = D_C f$ of $f$  belongs to $W^5(\mathcal{R}_B) \subset L^{\infty}(\mathcal{R}_B)$. We then can write
$$f - Z \tau D f = Z (I- \tau) D f. $$ 
Combining this observation with the bounds given by \eqref{Z1} for the norms of $Z$, $(I- \tau)$ and $D$, we obtain,  for all $f \in W^5(B)$,
\begin{eqnarray} \label{Z4}
\norm{f -  Z \tau D f }_{L^{\infty}(B)}   &=& \norm{Z (I-  \tau) D  f} _{L^{\infty}(B)} \nonumber \\
& \leq &  
 \frac{c^3}{N(\epsilon)} \norm{f}_{W^5(B)}. 
\end{eqnarray}
Since $D = D_C + Y_C$, for any  $f \in W^5(B)$ we have 
\begin{equation} \label{Z5}
f- Z \tau D f  = f - Z \tau Y_C f   -  Z \tau D_C f  = (I-U) f  - Z \tau h.
\end{equation}
From  \eqref{Z4} and \eqref{Z5} we then get that for all $f \in W^5(B)$  
$$
\norm{ (I- U)  f - Z \tau h  }_{L^{\infty}(B)} \leq \frac{c^3}{N(\epsilon)} \norm{f}_{W^5(B) }
$$
and, hence, 
since $\norm{ (I-U)^{-1} }_{L^{\infty}(B)} \leq 10$, we conclude that 
\begin{eqnarray} \label{Z6}
\norm{ f - (I-U)^{-1} Z \tau h  }_{L^{\infty}(B)} &\leq& 
\norm{  (I-U)^{-1} }_{L^{\infty}(B)} \norm{ (I- U)  f - Z \tau h  }_{L^{\infty}(B)} \nonumber \\ &\leq& 
10 \frac{c^3}{N(\epsilon)} \norm{ f }_{W^5(B) }.
\end{eqnarray}
\db{For  any  $f \in W^5(B)$, the expression of $Z_C h = Z_C D_C f$ given by equation \eqref{A2} then implies 
$$
f  - Z_C D_C f   =  f  - Z_C h = f - (I-U)^{-1} Z \tau h.     
$$
Hence equation \eqref{Z6} yields
$$
\norm{ f  - Z_C D_C f  }_{L^{\infty}(B)} \leq  10 \frac{c^3}{N(\epsilon)} \norm{f}|_{W^5(B) }.
$$
Our choice of $N$ in \eqref{Neps} forces $10 \frac{c^3}{N(\epsilon)}  \equiv \epsilon$. Thus,  for all  $f \in W^5(B)$ and all regions $C \subset B$ verifying $\rad(B) - \rad(C)  \leq  \eta(\epsilon)$, we get 
\begin{equation} \label{ZCXC}
\norm{(I  - Z_C D_C) f }_{L^{\infty}(B)} \leq  \epsilon  \norm{f}_{W^5(B)}.
\end{equation}
By definition \ref{eps.inverse}, $Z_C$ is thus an  $\epsilon$-accurate inverse of $D_C$. This  completes the proof of Theorem \ref{ROIalgo}. \qed
}

\paragraph{Proof of Corollary \ref{variantalgo}}

This proof is similar to the argument used in the proof above and will just be sketched. Select two $C^4$ approximations of the identity $(\sigma_N)$  in  $L^2(B)$  and $(\tau_N)$  in $L^2(\mathcal{R}_B)$ and set $\widetilde U_N = \sigma_N Z \tau_N Y_C$. By applying Lemma~\ref{contraction} to $\widetilde U_N$ we have that, given any $\epsilon >0$, there is $N = N(\epsilon)$ large enough and $\eta(\epsilon)$ small enough to ensure that, for all regions $C$ verifying $\rad(B) - \rad(C)  \leq  \eta(\epsilon)$, the operator $\widetilde U= \widetilde U_N$  satisfies $\norm{\widetilde U}_{L^{\infty}(B)}<0.9$. The iterative sequence $f_{j+1} = f_0 + \widetilde U f_j$ initialized by the new $f_0 = \sigma_N Z \tau_N g $ will then converge to a limit $Z_C g$ in $L^\infty(B)$. As in the argument above, this defines an $\epsilon$-accurate inverse $Z_C$ of $D_C$. \qed

\subsection{Extensions}
 Theorem~\ref{ROIalgo} and Corollary~\ref{variantalgo} can be easily  extended to the case where the set of sources $\Gamma$ is a sphere strictly including  $B$ rather than a curve, using arguments nearly identical to those used above. The constant $c$ in  \eqref{YC} then depends on the surface of the sphere $\Gamma$ instead of the length of the curve $\Gamma$. 
 
 Our ROI-reconstruction algorithm to approximately invert the $ROI$-truncated cone beam transforms $D_C$ can also be extended using nearly identical arguments to  generate approximate inverses of the $ROI$-truncated ray transforms~$X_C$.

\medskip
For discrete data $D_C f$ generated by a numerical $ROI$-truncated  cone beam transform of the unknown density $f$, our $ROI$ reconstruction algorithm \eqref{d:algo} can formally be applied, starting with any  discretized reconstruction algorithm $Z$ specific to the acquisition geometry at hand and known to perform well on  {\it non-truncated} data. From a formal point of view,  $Z$ can even be a `black-box' inversion software dedicated to inversion of non truncated data. Of course the proofs Theorem \ref{ROIalgo} do not a priori cover such cases from a theoretical point of view. Yet, our extensive numerical tests in Section~\ref{sec.num} indicate that our ROI-reconstruction approach does perform well in many practical CT setups.

\section{Numerical experiments} \label{sec.num}

In this section, we present extensive numerical experiments to evaluate the performance of our ROI reconstruction algorithm in multiple discrete settings with sources located on a smooth curve or a sphere. Since we had no access to ROI-truncated data acquired with an actual CT device enabling cone-beam ROI truncation, our numerical experiments simulated ROI-truncated cone-beam acquisition. We used four classical acquisition geometries and multiple spherical ROIs with different locations and sizes applied to several 3D density data.  The goal of these numerical experiments was to numerically quantify accuracy of our ROI reconstruction algorithm \textit{within the ROI} and to investigate how the ROI radius impacts this measure of accuracy. 

For a given cone-beam acquisition setup with target ball $B \subset \R^3$, a point $z$ in $B$ and a number $\nu > 0$, our Theorem \ref{ROIalgo} and Corollary \ref{variantalgo} imply the existence of a critical radius $\rho$ such that, for any spherical region $C \subset B$ with center $z$ and radius $\rad(C) > \rho$, and for any density $f$ with $\norm{f}_{L^{\infty}(B)} \leq \nu$, our ROI reconstruction algorithm from truncated data  will converge {\it within $C$} to a good approximation of the unknown $f$. Our numerical experiments provide practical evaluations of this {\it critical radius} $\rho$.

\subsection{Simulations of ROI-truncated acquisition}
We have simulated ROI-truncated cone-beam acquisition for three discretized densities $f$: a 3D Shepp-Logan phantom; a 3D scan of mouse tissue; a 3D scan of a human jaw. For each discretized density $f$, given by a  3D image of size   $256^3$ voxels,  we  have first computed discrete non trucated cone-beam projections $D f$ by simulating discrete acquisition and used these data to generate ROI-truncated 3D projections for four distinct acquisition geometries
with the following parameters: 
\begin{enumerate}
\item[(i)] {\it Spherical tomography with sources on a full spherical surface.}   Ray discretization: 3 degrees in the polar direction, 5 degrees in the azimuthal direction; scanning radius: 400 voxels; number of detector rows: 256; source-detector distance = 900 voxels. 

\item[(ii)] {\it Spiral tomography with sources on a helix. } 
Helical pitch: 35 voxels; 8 turns to scan the whole object; number of source positions: 128 per complete turn; scanning radius = 384 voxels; number of detector rows = 16; source-detector distance: 768 voxels. 
\item[(iii)] {\it C-arm tomography with sources on a circle.} Scanning radius: 1472 voxels; number of source positions: 360; detector size: 256 rows, 256 columns; detector spacing: 1 voxel; source-detector distance: 1472 voxels. 
\item[(iv)] {\it Twin circles tomography with sources on two concentric circles, contained in orthogonal planes on $\R^3$.} Common radius of the two circles: 1472 voxels; number of source positions: 360 per circle; detector size: 256 rows, 256 columns; detector spacing: 1 voxel; source-detector distance: 1472 voxels.
\end{enumerate}

For each one of these four acquisition setups, we selected four concentric spherical ROI $C$ with radius values (in voxels)  equal to  45, 60, 75, 90. Each such ROI $C$ was used to truncate the discretized projection data $Y=D f$ to the rays intersecting $C$ and thus to generate a discretized version of the ROI-truncated data $Y_C =D_C f$. Note that non truncation corresponded to a much larger spherical radius (221 voxels) covering the entire 3D density volume.

\subsection{Numerical implementations of our ROI reconstruction algorithm.}

For each 3D discrete density function $f$, each cone-beam acquisition setup and each spherical ROI $C$, we have implemented our iterative ROI reconstruction algorithm to compute a reconstruction  $Z_C f$ of $f$ using only the ROI-truncated data $D_C f$. According to our general scheme~\eqref{d:algo} and Corollary~\ref{variantalgo}, we apply the iterative ROI reconstruction formula $f_{j+1}= f_0 + U f_j$ where $U = \sigma  Z  (D - D_C)$,  $\sigma$
is a regularization operators and $Z$ is the inverse of the {\it non-truncated} cone-beam transform implemented using the following specific methods, according to the acquisition setup. 

For each one of our simulated cone-beam acquisition setups, the inverse $Z$ of the {\it non-truncated } cone-beam transform was implemented as follows:
\begin{enumerate}
\item[(i)] Spherical tomography: Inversion of  {\it non-truncated} cone-beam transform by filtered back-projection (FBP) \cite{natterer:imagerec}.
\item[(ii)] Spiral tomography: Inversion of  {\it non-truncated} cone-beam transform by Katsevich's inversion formula  \cite{Katsevich:spiralCT04}. 
\item[(iii)] C-arm tomography: Inversion of {\it non-truncated} cone-beam transform by a well known FDK algorithm  \cite{feldkamp1984practical}. 
\item[(iv)] Two circles tomography: Inversion of {\it non-truncated} cone-beam transform by a discretized version of Grangeat's formula  outlined in  Defrise and Clack \cite{defrise1994cone}.  
\end{enumerate}

\paragraph{Choice of  a regularization operator $\sigma$ on the euclidean ball $B$.} For all our acquisition setups, we  used similar  discretized versions of the regularization operator $\sigma: L^2(B) \to W^4(B)$ based on {\it  wavelet thresholding}. That is, to compute $\sigma h $ for any $h$ in $L^2(B)$, we first expanded $h$ using standard Daubechies wavelets Daub4 \cite{Mallat:2008}  in $\R^3$ to generate the wavelet  decomposition of $h$
$$
h = \sum_{m, n, i}  a(m,n,i) \phi_{m,n,i}.
$$
Each  wavelet $\phi_{m,n,i}$  in this family is a  $C^4$ function  indexed by the discretized position $(m,n)$ of its compact support and by an integer scale parameter $i \ge 0$.  At the coarsest scale $i = 0$, no truncation or shrinkage was applied to the wavelets coefficients $a(m,n,i)$. At finer scales $i \geq 1$,  the wavelets coefficients $ a(m,n,i) $ were set to zero whenever 
$| a(m,n,i) | < THR_i$. where the thresholds $THR_i$ were selected to discard  90\% of wavelet coefficients. \db{We found that the performance of the algorithm is not very sensitive to the choice of the percentage of discarded wavelet coefficients, that is, performance would remain essentially the same by discarding 70-90\% of wavelet coefficients. Additional details can be found in~\cite{Sen_thesis}.}

The new wavelet expansion generated by this coefficients truncation defined the function $\sigma h$ which obviously belonged to $W^4(B)$. This operator $\sigma$ is  non linear but can be well approximated by linearized versions which implement smooth shrinkage of the wavelets coefficients instead of abrupt truncation (see \cite{Sen_thesis}.) 

\paragraph{Stopping rules for our ROI reconstruction} Let $C$ be the spherical ROI. As a stopping criterion, we adopted a standard rule so that the algorithm \eqref{d:algo} is to stop the iteration over the index $j$ when $f_j$ and $f_{j+1}$ become  close enough {\it within $C$}; in particular, as long as
$$
\norm{f_{j+1} - f_j}_{L^1(C)} \leq b  
$$
for some small tolerance $b$, e.g.  $b = 0.02$. We automatically stop the ROI iterative reconstruction at $j=40$ to avoid unnecessary computation as we found that, for all our numerical experiments, as soon as the radius of $C$ was slightly superior to a critical radius, 40 iterations were amply sufficient to achieve convergence.

\subsection{Performances of numerical ROI reconstructions}

For each one of our three densities $f$,  each one of our  four x-ray acquisition setups,  and each one of our selected spherical ROIs $C$, our simulations generated a discrete version $g = D_C f$ of the ROI-truncated cone-beam projections of $f$. Then the numerical application  of our iterative ROI reconstruction algorithm to these truncated data $g$ provided a discretized  approximation $Z_C f $  of  the  "unknown" $f$. To assess the accuracy of our  discretized ROI reconstruction $Z_C f$, we have evaluated an  \textit{  ROI  Relative $L^1$ Error }of Reconstruction   within $C$ defined by the following ratio $RL_1$ of two  discretized $L^1(C)$ norms
\begin{equation*}\label{error}
RL_1 = \frac{ \norm{f - Z_C f }_{L^1(C)} } {\norm{f}_{L^1(C)}}
\end{equation*}
{For each one of the 48 ROI reconstruction cases indicated above, we have recorded the Relative $L^1$ reconstruction error computed within the ROI in Table \ref{accuracytable}. As indicated above, our iterative algorithm \eqref{d:algo} uses different numerical routines to implement the non-truncated inverse operator $Z$ depending on the acquisitions geometry. Namely, in the case of sources on a sphere, $Z$ is implemented using the FBP algorithm; for sources on a spiral curve, $Z$ is implemented using the Katsevich's inversion formula; for sources on a circular curve, $Z$ is implemented using the FDK algorithm; for sources on a twin-circle curve, $Z$ is implemented using a version of Grangeat's formula. The number of iterations needed to achieve convergence of our ROI reconstruction algorithm was bounded above by 40 but the algorithm was found to converge (according to the stopping rule given above) with a much smaller number of iterations, typically between 10-12 iterations for sources on a curve.}

\begin{table} 
\centering
\begin{tabular}{|c|c|c|c|c|c|}
\hline
& & \multicolumn{4} {|c|}{\textbf{Sources locations}}  \\ 
 \cline{3-6}
\textbf{Density} & \textbf{ROI} & Spherical & Spiral & Circle & Twin circles  \\ 
\textbf{data} &\textbf{radius} &  &  &  &  \\ \hline 
\multirow{4}{*}{Shepp-Logan} 
&  45 vox & 10.3\% & 10.9\% & 13.2\% & 14.8\% \\ 
&  60 vox & 8.6\% & 9.1\% & 11.6\% & 14.7\% \\ 
&  75 vox & 7.6\% & 8.3\% & 7.4\% & 8.9\% \\ 
&  90 vox & 7.3\% & 8.0\% & 4.4\% & 4.8\% \\ \hline
\multirow{4}{*}{Mouse tissue} 
&  45 vox & 10.8\% & 11.4\% & 11.6\% & 12.5\% \\ 
&  60 vox & 8.8\% & 9.7\% & 11.1\% & 9.4\% \\ 
&  75 vox & 7.9\% & 8.8\% & 8.4\% & 8.3\% \\ 
& 90 vox & 7.5\% & 8.4\% & 7.1\% & 7.8\% \\ \hline
\multirow{4}{*}{Human jaw} 
&  45 vox & 11.4\% & 11.9\% & 12.9\% & 15.0\% \\ 
&  60 vox & 9.6\% & 10.8\% & 12.8\% & 13.3\% \\ 
&  75 vox & 9.0\% & 9.7\% & 10.2\% & 10.2\% \\ 
&  90 vox & 8.2\% & 8.5\% & 9.8\% & 9.8\% \\ \hline
\end{tabular}
\caption{Relative $L^1$ error of ROI reconstruction. The table shows the reconstruction accuracy within the ROI using four ROI radii for three 3D density data and four cone-beam acquisition geometries. Each density data set has size $256^3$.}\label{accuracytable}
\end{table}

For each one of the various combinations of density data and acquisition setup, these accuracy results yield  an estimate of the critical ROI  radius $\rho$  enabling a relative  ROI reconstruction accuracy inferior or equal  to $10\%$. Critical radius estimates are displayed in Table~\ref{criticaltable}.

\begin{table}
\centering 
\begin{tabular}{|c|c|c|c|c|}
\hline
&  \multicolumn{4} {|c|}{\textbf{Source locations}}  \\ 
 \cline{2-5}
{\bf Density data} &   Spherical & Spiral & Circle & Twin circles \\ \hline 
\multirow{1}{*}{Shepp-Logan} 
&   52 vox & 56 vox & 67 vox & 73 vox \\ \hline
\multirow{1}{*}{Mouse tissue} 
&   52 vox & 57 vox & 66 vox & 49 vox \\ \hline
\multirow{1}{*}{Human jaw} 
&  57 vox & 70 vox & 82 vox & 82 vox \\ \hline
\end{tabular}
\caption{Critical radius of convergence. For three 3D density data and four cone-beam acquisition geometries, the table shows the critical ROI radius above which the relative accuracy of the iterative ROI reconstruction was found to be less than 0.1.  }  \label{criticaltable}
\end{table}

The best performances of our ROI reconstruction  algorithm naturally occur for spherical acquisition. Indeed for the somewhat academic spherical setup, the number of projections available is  much larger than for the three other setups where sources are located on a curve.

For the twelve situations evaluated here, we obtain a range from 52 to 82 voxels for the critical radius $\rho$ yielding a 10\% accuracy in ROI reconstruction. This compares very favourably to the maximal ROI radius corresponding  to non truncation. Indeed when one goes from non truncation to a critical spherical ROI, the reduction in irradiated volume ranges from  70\% to 98\%, indicating a quite strong ``formal" reduction in x-ray exposure, while the loss in relative reconstruction accuracy  is only of the order of 7\%.

Note also that the actual critical radius estimates obtained here by simulations are much smaller that the theoretical upper  bounds  used in the proof of Theorem~\ref{ROIalgo}.

For a fixed ROI radius, the ROI Relative Reconstruction error is lower for the {\it Shepp-Logan phantom}  than for {\it Mouse Tissue} or  {\it Human Jaw} density data. Indeed, when the ROI radius is larger than the critical radius, our iterative ROI reconstruction essentially converges within the ROI to a regularization $\sigma f$ of $f$. The ROI reconstruction error in $L^1(C)$ can roughly be viewed as the sum of two terms, a `convergence' error $\norm{Z_C f - \sigma f}_{L^1(C)}$ and a `regularization' error $\norm{f - \sigma f}_{L^1(C)}$. To highlight the regularization effect, we have computed the relative `regularization error' within $C$ given by 
$$
\frac{\norm{f - \sigma f}_{L^1(C)}}{ \norm{f}_{L^1(C)}}
$$
For an ROI radius of 70 voxels, this regularization error is equal to $1.1\%$ for the 3D Shepp-Logan phantom, and to $2.4\%$ for the Mouse Tissue and Human Jaw 3D data, because the piecewise constant Shepp-Logan phantom density can be approximated by our wavelet-based regularization operator much more effectively than the more textured Mouse Tissue and Human Jaw densities. So for the Human Jaw data the  regularization error contributes about half of the relative $L^1$ reconstruction error.

\begin{figure*}
\centering
\includegraphics[height=4.1in]{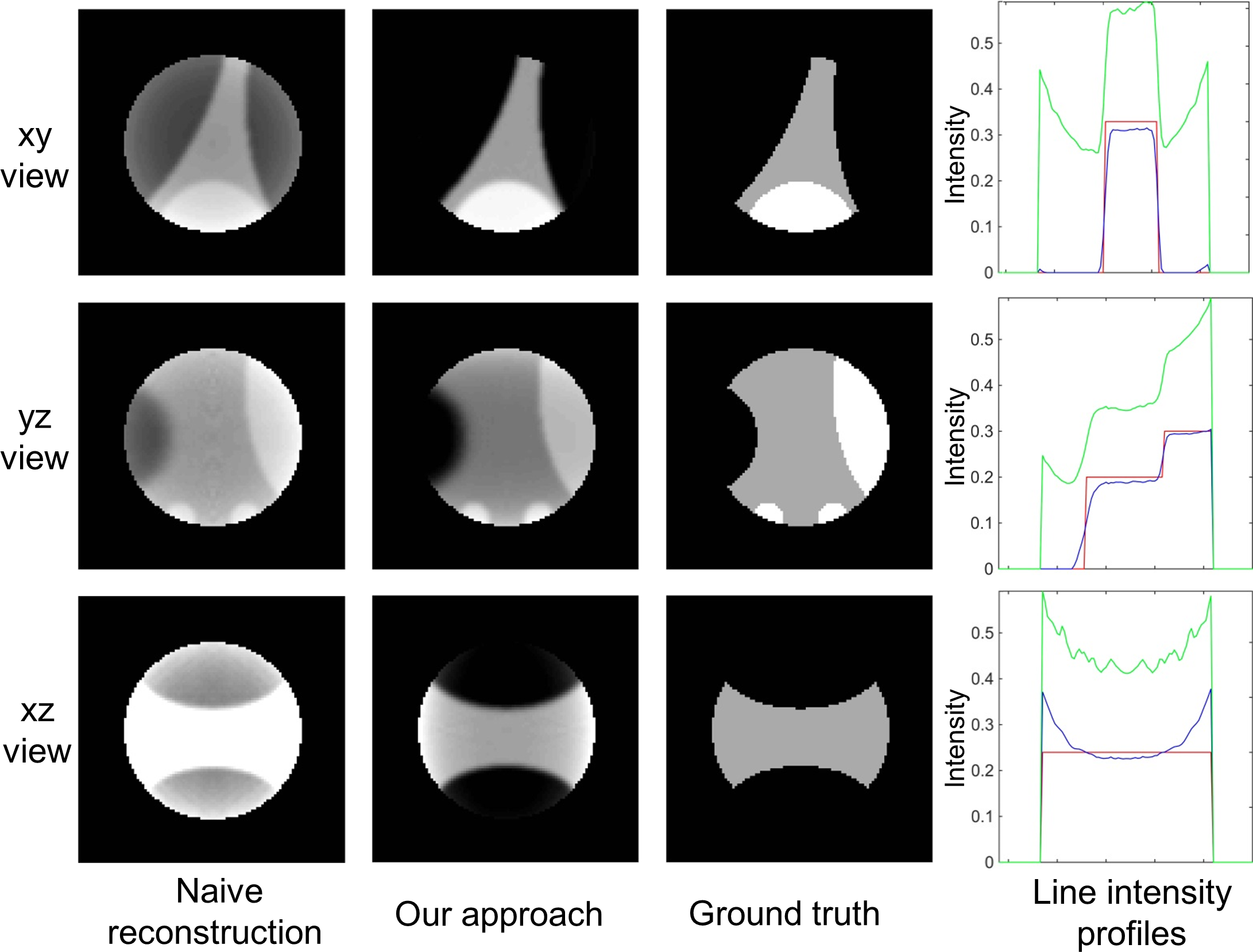}
\caption{Visual comparison of ROI reconstruction for 3D Shepp-Logan phantom
using simulated Twin Circles acquisition and truncation of projection data. ROI radius = 45 voxels. Middles sections are shown from the $xy$, $yz$ and $xz$ planes. From left to right:
inversion by one-step Grangeat formula; our iterative ROI reconstruction; ground truth. The last column shows intensity profiles corresponding to the middle row of the images. Green: one-step Grangeat formula; blue: our algorithm; red: ground truth. }
\label{twincircles_view_shepp}
\end{figure*}

\begin{figure*}
\centering
\includegraphics[height= 4.05 in]{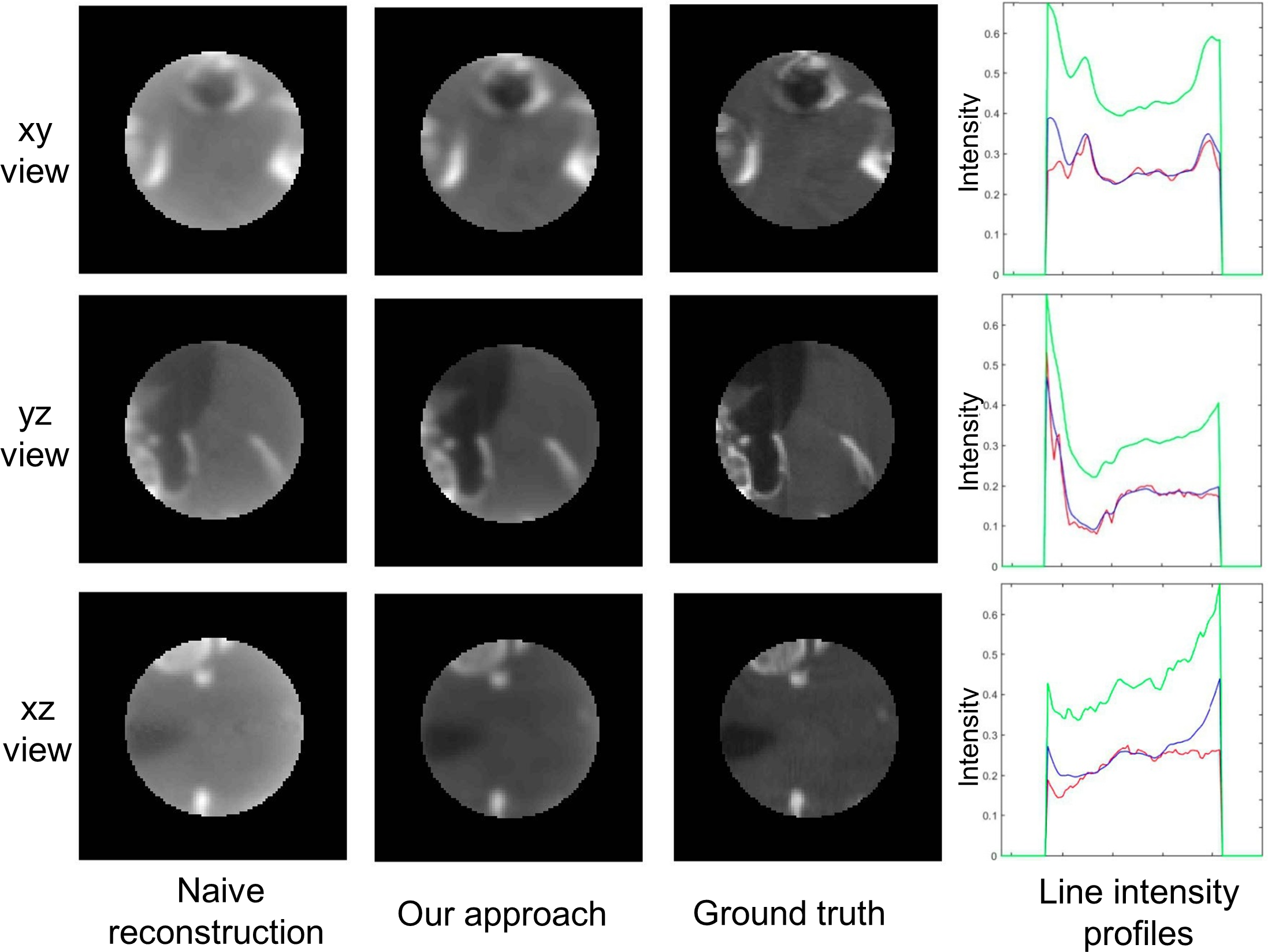}
\caption{Visual comparison of ROI reconstruction for Mouse Tissue data
using simulated Twin Circles acquisition and truncation of projection data. ROI radius = 45 voxels.
Middles sections are shown from the $xy$, $yz$ and $xz$ planes. From left to right:
inversion by one-step Grangeat's  formula; our iterative ROI reconstruction; ground truth. The last column shows intensity profiles corresponding to the middle row of the images. Green: one-step Grangeat formula; blue: our algorithm; red: ground truth.}
\label{rec_view_mouse}
\end{figure*}

\db{To illustrate visually the overall performance of our iterative ROI reconstruction from truncated cone-beam data, we have include several examples. 
For all these examples, the size of the images is $256^3$ voxels and the ROI radius is 45 voxels.
In Figures~\ref{twincircles_view_shepp}-\ref{rec_view_mouse} we show horizontal, coronal, and sagittal planes from our 3D reconstruction of the Shepp-Logan 3D Phantom and Mouse Tissue data using simulated Twin Circle acquisition. We also include line profiles to compare our reconstruction against ground truth and one-step inversion formula. }

\begin{figure*}
\includegraphics[height=3.3 in]{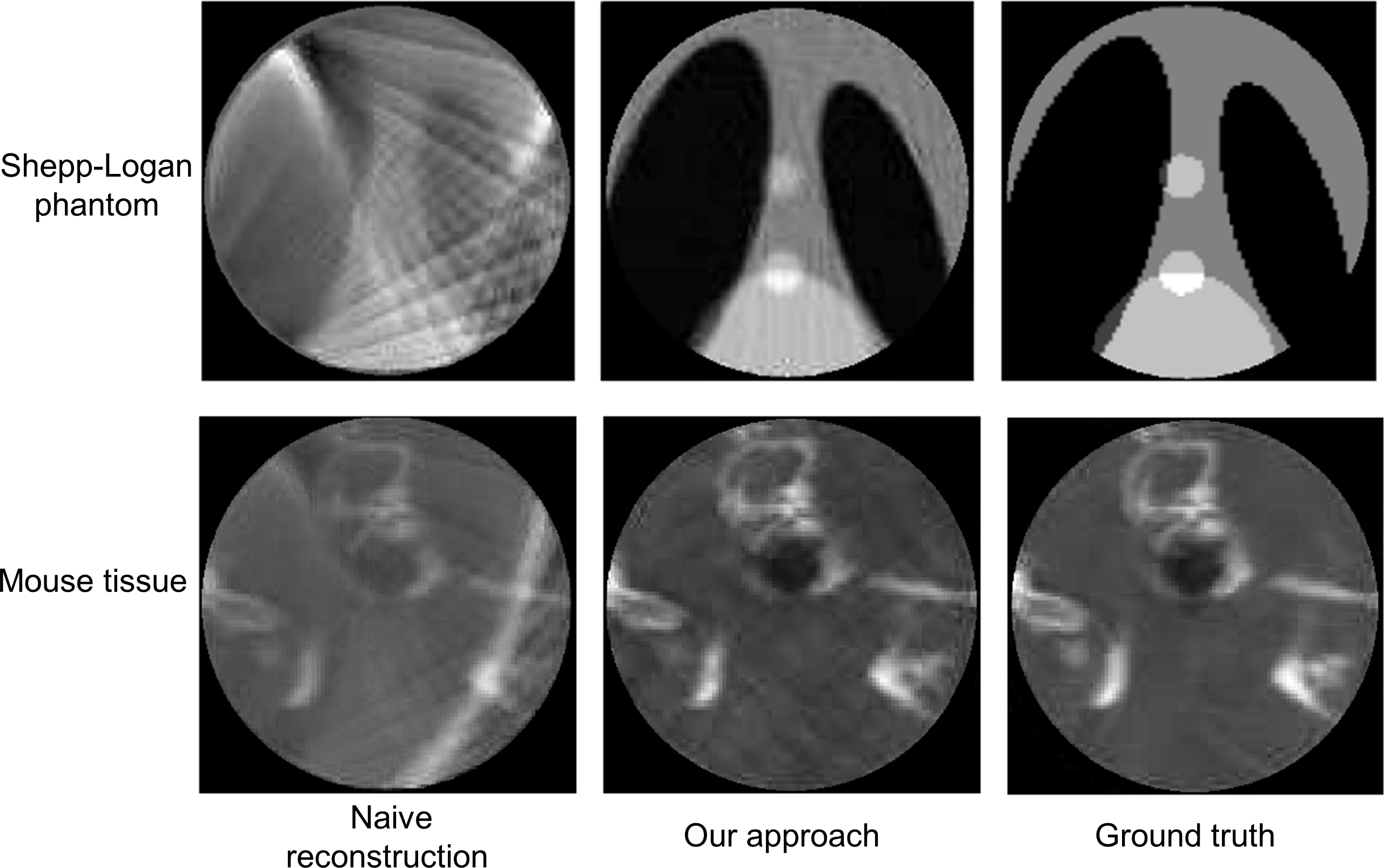}
\caption{Visual comparison of ROI reconstruction for 3D Shepp-Logan phantom and mouse tissue using simulated spiral acquisition and truncation of projection data. A
representative horizontal section from the 3D reconstructed volume is shown. From left to right:
inversion by one-step Katsevich formula; our iterative ROI reconstruction; ground truth.}
\label{spiral}
\end{figure*}

\begin{figure*}
\includegraphics[height= 3.3 in]{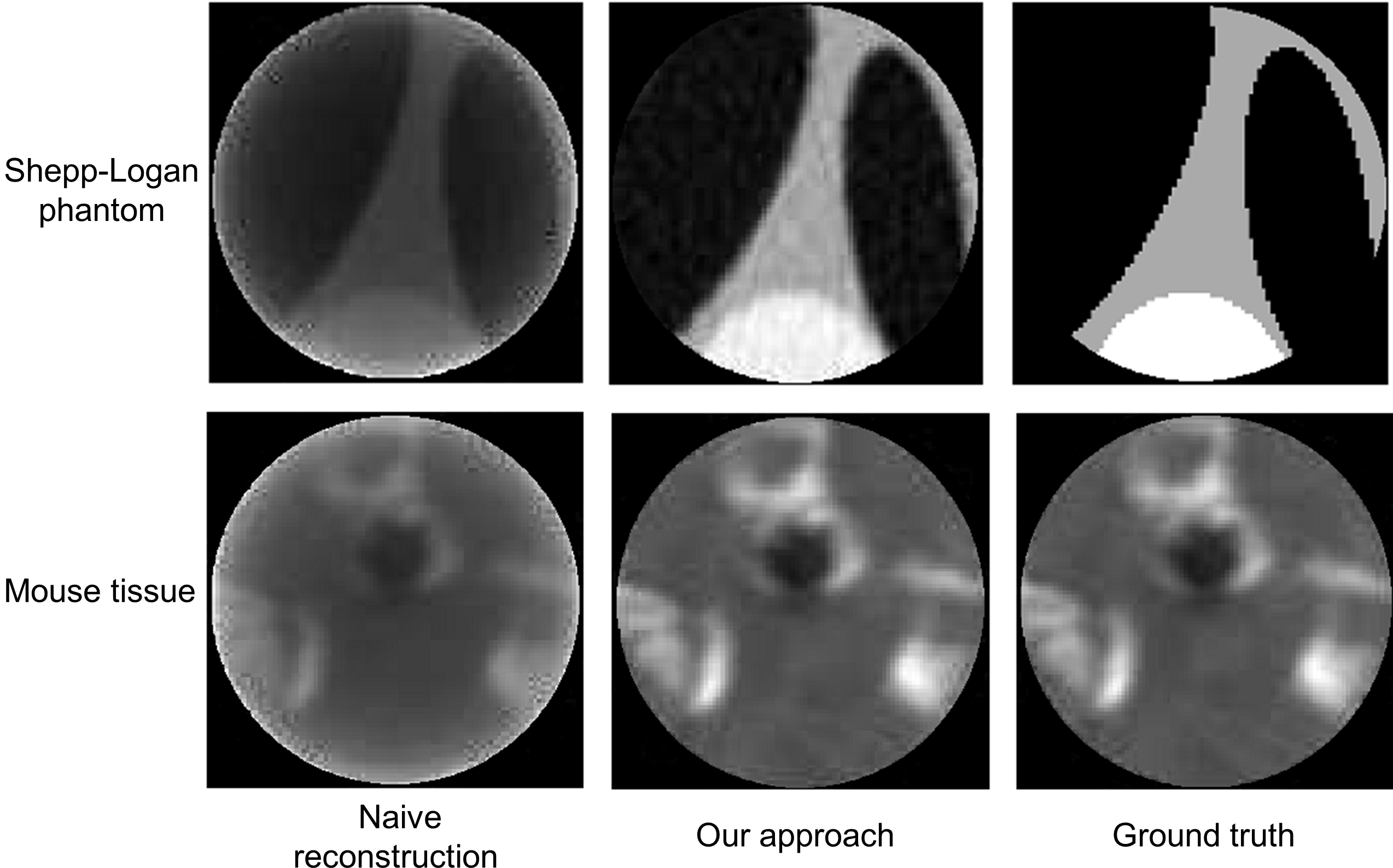}
\caption{Visual comparison of ROI reconstruction for 3D Shepp-Logan phantom and mouse tissue
using simulated C-arm acquisition and truncation of projection data. A
representative horizontal section from the 3D reconstructed volume is shown. From left to right:
inversion by one-step FDK algorithm; our iterative ROI reconstruction; ground truth.} 
\label{circular}
\end{figure*}

\db{In Figures~\ref{spiral}-\ref{circular}, we show horizontal sections from the reconstructed volumes of the Shepp-Logan 3D Phantom and Mouse Tissue data using 
simulated spiral and C-arm acquisitions.}

In all these figures, the comparison of results from our iterative ROI reconstructions with those obtained by the classical one-step inversion formulas originally devised for reconstruction from {\it non-truncated} cone-beam data show that, as expected, the one-step inversion formulas for non-truncated data perform poorly when applied to ROI-truncated cone-beam data, and display multiple visual artifacts especially near the ROI boundary. By contrast, our iterative ROI reconstruction results are very satisfactory even for relatively small ROI radii.
\db{Compared to the ground truth, our ROI reconstruction shows some blurring which is due to the wavelet-based regularization step. Since our wavelet filters have finite support and length 4 pixels, they have a rather limited impact on spatial resolution. The blurring effect is consistent with our theoretical prediction since our algorithm generates an approximation of the exact solution which is a smoother version of the true image.
}

\db{Note also that the images reconstructed using one-step Katsevich formulas in Figure~\ref{spiral} exhibit streak artifacts. This is a common and known issue in spiral tomography, cf. \cite{yazdi2008}. Our regularized reconstruction significantly reduces these artifacts  through the wavelet-based regularization.
}

\section {Conclusion } \label{s.conclu}
In this paper, we have examined the problem of ROI tomographic reconstruction using truncated cone-beam data, a problem of high relevance in many applications. For both our theoretical and numerical analysis, we considered fairly generic cone-beam acquisition setups, with sources located on arbitrary bounded smooth curves $\Gamma$ in $\R^3$ verifying classical Tuy's condition.  In all these cases,  it is known that the  {\it non-trucated} cone-beam transform $D f$ of smooth densities $f$ admits an explicit inverse $Z$ but $Z$ cannot directly reconstruct $f$ from ROI-truncated data. 

To deal with the reconstruction from ROI-truncated data, we have developed and rigorously analyzed a new iterative ROI reconstruction method valid for densities $f$ in $L^{\infty}(B)$, where $B$ is a bounded ball in $\R^3$, which iterates  a linear contraction endomorphism $U$ of  $L^{\infty}(B)$. The operator $U$  is constructed by combining forward ROI-truncated projections, backward inversion by the  operator $Z$ and appropriate regularization operators defined in image and/or projection space. Our main theoretical result is that, given $\epsilon >0$, for   spherical regions of interest $C\subset B$ with radius larger than a critical radius $\rho(\epsilon)$: (i) our iterative ROI reconstruction from ROI-truncated data converges in $L^{\infty}(B)$ to a density estimate $\hat{f}$ such that  $\norm{ \hat{f} - f }_{\infty }  \leq \epsilon \, \norm{ f }_{\infty } $; (ii) our iterative ROI reconstruction algorithm generates a bounded linear operator $Z_C: L^{\infty}(\mathcal{R}_B)  \to L^{\infty}(B)$, where $\mathcal{R}_B$ is the Riemannian manifold of all x-rays emitted by sources on a curve $\Gamma$ outside $B$. The operator $Z_C$ is an  $\epsilon$-inverse of the ROI-truncated cone-beam transform $D_C: L^{\infty}(B)  \to L^{\infty}(\mathcal{R}_B) $, that is $ \norm{ I - Z_C D_C }_{L^{\infty}(B)} < \epsilon$. These results also extend to the case of spherical acquisition in $\R^3$ and to the ray transform.

Even though iterative methods for ROI CT reconstruction already appeared in the literature, up to the knowledge of the authors, no theoretical result was known so far about the existence of a critical radius ensuring the convergence of an iterative ROI CT reconstruction scheme.  

We numerically verified our theoretical results using simulated 3D acquisition of ROI-truncated cone-beam data for four classical acquisition geometries (spherical, spiral, circular arm, twin orthogonal circles), using three different density functions and multiple ROI radii and locations. All numerical experiments show that,  for $\epsilon$ moderately small, e.g., $\epsilon = 0.1$, the critical ROI radius $\rho(\epsilon)$ is fairly small with respect to the support of the density function.

\section*{Acknowledgements}
Authors thank M. Motamedi and I. Patrikeev,
at the Center of Biomedical Engineering, UTMB, for providing the micro-CT images of the Mouse tissue. A.S. and R.A. acknowledge support by a
Methodist Hospital grant provided by Dr. K. Li, Chair of Radiology.
B.G.B. acknowledges partial support by NSF DMS 1412524 and by the Alexander von Humboldt foundation, and for the great hospitality in G. Kutyniok's group at the Technische Universit{\"at} Berlin, where part of this work was completed.
D.L. acknowledges partial support by NSF DMS 1008900 and 1320910.

\begin{appendix}
\section*{Appendix: Sobolev imbeddings}   \label{app.sobolevimbed}

In Section~\ref{smoothspaces}, to show the regularity of the linear operator $Z$, we make use of Sobolev imbedding theorems. We quote a special case of a result by Aubin (Theorem 2.34 in~\cite{aubin_82}) in this context.

\begin{theorem}[\cite{aubin_82}]
If $\overline{\mathcal M}$ is a compact Riemannian manifold of dimension $n$ with $C^1$-boundary and interior $\mathcal M$, then 
$$
 W^k({\mathcal M} ) \subset C^\alpha(\overline{\mathcal M})
$$ 
and this imbedding is compact if $k-\alpha>n/2$.
\end{theorem}

By the compactness of the embedding, it is also continuous, meaning that if $k-\alpha>n/2$ then there exists $c>0$ such that for each $f \in W^4({\mathcal M})$,
$$
  \| f\|_{C^\alpha(\overline{\mathcal M})} \le c \| f \|_{W^4({\mathcal M})}  \, .
$$ 
In particular, if $n=3$ and $k=4$, then we can choose $\alpha=2$ and have the following result.

\begin{corollary}  
If $\overline{\mathcal M}$ is a compact Riemannian manifold of dimension $n$ with $C^1$-boundary and interior $\mathcal M$ then 
$$
  \| f\|_{C^2(\overline{\mathcal M})} \le c \| f \|_{W^4({\mathcal M})}  \, .
$$ 
\end{corollary}

\end{appendix}

\bibliographystyle{IEEEtran}
\bibliography{autobib}

\end{document}